\documentclass[pdflatex,sn-mathphys-num]{sn-jnl}

\usepackage{graphicx}%
\usepackage{subcaption}%
\usepackage{siunitx}%
\usepackage{physics}%
\usepackage{multirow}%
\usepackage{amsmath,amssymb,amsfonts}%
\usepackage{amsthm}%
\usepackage{mathrsfs}%
\usepackage[title]{appendix}%
\usepackage{xcolor}%
\usepackage{textcomp}%
\usepackage{manyfoot}%
\usepackage{booktabs}%
\usepackage{algorithm}%
\usepackage{algorithmicx}%
\usepackage{algpseudocode}%
\usepackage{listings}%
\usepackage{setspace}
\usepackage{lineno}
\usepackage[normalem]{ulem}

\theoremstyle{thmstyleone}%
\theoremstyle{thmstyletwo}%

\theoremstyle{thmstylethree}%

\begin{document}


\title[Article Title]{Gate-Tunable Giant Negative Magnetoresistance in Tellurene Driven by Quantum Geometry}


\author*[1]{\fnm{Marcello B.} \sur{Silva Neto}}\email{mbsn@if.ufrj.br}
\equalcont{These co-first-authors contributed equally to this work.}

\author[2,3]{\fnm{Chang} \sur{Niu}}\email{niu43@purdue.edu}
\equalcont{These co-first-authors contributed equally to this work.}

\author[4]{\fnm{Marcus V. O.} \sur{Moutinho}}\email{moutinho@xerem.ufrj.br}

\author[5]{\fnm{Pierpaolo} \sur{Fontana}}\email{pierpaolo.fontana@uab.cat}

\author[6]{\fnm{Claudio} \sur{Iacovelli}}\email{claudio.iacovelli@hotmail.it}

\author[7]{\fnm{Victor} \sur{Velasco}}\email{vvelasco@sissa.it}

\author[1]{\fnm{Caio} \sur{Lewenkopf}}\email{lewenkopf@if.ufrj.br}

\author*[2,3]{\fnm{Peide D.} \sur{Ye}}\email{yep@purdue.edu}

\affil[1]{\orgdiv{Instituto de F\'isica}, \orgname{Universidade Federal do Rio de Janeiro}, \orgaddress{\city{Rio de janeiro}, \postcode{21941-972}, \state{RJ}, \country{Brazil}}}

\affil[2]{\orgdiv{Elmore Family School of Electrical and Computer Engineering}, \orgname{Purdue University}, \orgaddress{\city{West Lafayette}, \postcode{47907}, \state{Indiana}, \country{United States}}}

\affil[3]{\orgdiv{Birck Nanotechnology Center}, \orgname{Purdue University}, \orgaddress{\city{West Lafayette}, \postcode{47907}, \state{Indiana}, \country{United States}}}

\affil[4]{\orgname{Universidade Federal do Rio de Janeiro - Campus Duque de Caxias}, \orgaddress{\city{Duque de Caxias}, \postcode{25240-005}, \state{RJ}, \country{Brazil}}}

\affil[5]{\orgdiv{Departament de F\'isica}, \orgname{Universitat Aut\`onoma de Barcelona}, \postcode{08193}, \orgaddress{\city{Bellaterra}, \country{Spain}}}

\affil[6]{\orgdiv{Independent Researcher}, \orgaddress{\city{Barcelona}, \postcode{08193}, \country{Spain}}}

\affil[7]{\orgname{International School for Advanced Studies (SISSA)}, \orgaddress{\city{Trieste}, \postcode{I-34136}, \country{Italy}}}

\doublespacing


\abstract{
Negative magnetoresistance in conventional two-dimensional electron gases is a well-known phenomenon, but its origin in complex and topological materials, especially those endowed with quantum geometry, remains largely elusive. 
Here, we report the discovery of a giant negative magnetoresistance, reaching a remarkable $- 90\%$ of the resistance at zero magnetic field, $R_0$, in $n$-type tellurene films. 
This record-breaking effect persists over a wide magnetic field range (measured up to $35$ T) at cryogenic temperatures and is suppressed when the chemical potential shifts away from the Weyl node in the conduction band, strongly suggesting a quantum geometric origin. 
We propose two novel mechanisms for this phenomenon: a quantum geometric enhancement of diffusion and a magnetoelectric spin interaction that locks the spin of a Weyl fermion, in cyclotron motion under crossed electric $\boldsymbol{\cal E}$ and magnetic ${\bf B}$ fields, to its guiding-center drift, $(\boldsymbol{\cal E}\times{\bf B})\cdot\sigma$. We show that the time integral of the velocity auto-correlations promoted by the quantum metric between the spin-split conduction bands enhance diffusion, thereby reducing the resistance. This mechanism is  experimentally confirmed by its unique magnetoelectric dependence, $\Delta R_{zz}(\boldsymbol{\cal E},{\bf B})/R_0=-\beta_{g}(\boldsymbol{\cal E}\times{\bf B})^2$, with $\beta_{g}$ determined by the quantum metric.
Our findings establish a new, quantum geometric and non-Markovian memory effect in magnetotransport, paving the way for controlling electronic transport in complex and topological matter.
}




\maketitle

\section{Introduction}
\label{sec:introduction}

The discoveries of giant magnetoresistance (GMR) in metallic multilayers \cite{Baibich1988,Parkin1990} and of colossal magnetoresistance (CMR) in correlated oxides \cite{ColossalMR} illustrate how intricate electron interactions produce stark resistance changes. In bulk materials, these effects arise from spin-polarized scattering or orbital coupling \cite{Zutic2004}, while in thin films, reduced dimensionality unlocks  magnetoresistance via quantum confinement \cite{Novoselov2005} or topological protection \cite{Hasan2010}. The advent of two-dimensional (2D) materials has particularly transformed the landscape, offering atomic-scale thickness and electrostatic gate-control that reveal novel magnetoresistive behaviors inaccessible in conventional systems.

Rare in bulk systems, negative magnetoresistance (NMR) has been observed in various 2D materials with distinct magnitudes, field ranges, and underlying mechanisms. In graphene, for example, weak NMR of $-2\%$ at $10$ mT arises from weak localization (WL) due to quantum interference of backscattered electron waves \cite{Tikhonenko2008}. Monolayer MoS$_2$ exhibits stronger NMR of $-10\%$ at $5$ T attributed to intervalley scattering mediated by spin-orbit coupling \cite{Cui2015}, while transition metal dichalcogenide heterostructures (e.g., WS$_2$/WSe$_2$) show NMR of $-12\%$ at $3$ T, from moiré-induced Berry curvature \cite{Xu2025}. Beyond these, black phosphorus displays NMR of $-15\%$ at $2$ T due to anisotropic scattering in its puckered lattice \cite{Li2017}, and topological insulators, like Bi$_2$Se$_3$ thin films, show NMR up to $-20\%$ at $1–2$ T \cite{He2011} linked to helical surface states protected by time-reversal symmetry. Meanwhile, Cr$_2$S$_3$, a 2D ferromagnet, demonstrates NMR of $-25\%$ at $100$ K due to spin-filtering effects at magnetic domain walls \cite{Moinuddin2021}, and twisted bilayer graphene near magic angles exhibits NMR of $-30\%$ at $1$ T. Most remarkably, in Weyl semimetals (e.g., WTe$_2$ films), NMR reaches $-60\%$ at $60$ T under pulsed fields, driven by the chiral anomaly, in the ultraquantum limit \cite{Ali2014}.  

A non-saturating magnetoresistance signals a departure from simple Markovian transport \cite{Levitov2008}, a memory of the carrier's past history, encoded in its momentum or phase, influencing its future trajectory. For example, in systems with dilute, strong scatterers (e.g., hard disks or antidots of radius $a$), and at low fields ($\omega_c \tau \ll 1 $, where $\omega_c$ is the cyclotron frequency and $\tau$ is the relaxation time), the dominant memory effect is purely classical and geometric: backscattering events induce a NMR, $\Delta R_{xx}/R_0 \sim -(\omega_c \tau)^2 / \beta_0$, where $\beta_0 = a/\ell$, due to correlated returns to scatterers of mean-free-path $\ell$ \cite{Dmitriev2001}, as described by the Lorentz gas model \cite{Cheianov2003}. 
For 2DEGs with smooth disorder (e.g., remote impurities), guiding center diffusion leads to 
$\Delta R_{xx}/R_0 \sim -(\omega_c / \omega_0)^2$ at low fields ($\omega_c \ll \omega_0$), where $\omega_0 \sim v_F (a^2 \ell_S \ell_L)^{-1/4}$ \cite{Mirlin2001,Polyakov2001}, with $\ell_S$ and $\ell_L$ representing 
the correlation lengths of the small and large disorder potentials \cite{Mirlin2001,Fogler1997}.
Beyond these classical pictures, NMR can also emerge from the quantum mechanical memory of the electron's phase in disordered systems, from WL and electron-electron interactions. 
At low fields ($B \lesssim \hbar / e \ell_{\phi}^{2}$, where $\ell_{\phi}$ is the phase coherence length), WL induces NMR, $\Delta R_{xx}/R_0 \sim -1\%$, through constructive interference of time-reversed paths \cite{Altshuler1980}, becoming irrelevant in the classical regime ($\hbar \omega_{c} \ll k_{B} T$) or under strong inelastic scattering \cite{Gornyi2003}. 
In contrast, interactions drive NMR via diffusive ($k_{B} T \tau / \hbar \ll 1$) or ballistic ($k_{B} T \tau / \hbar \gg 1$) corrections, producing a parabolic field dependence, $\Delta R_{xx}/R_0 \sim -B^{2}$, with larger magnitudes ($\sim 1-10\%$) observed in clean, high-density GaAs heterostructures \cite{Paalanen1983,Li2003}. 

Here we report the observation of gate-controlled, giant negative magnetoresistance (GNMR) in $n$-type tellurium (Te) flakes, originating from its intrinsic (Weyl) quantum geometry and from the presence of a uniform, lone-pair-induced polarization electric field $\boldsymbol{\cal E}$ \cite{OurPRL2025}. 
Our GNMR reaches up to $-90\%$ at $32$ T, far exceeding the field range and magnitude of prior 2D systems. We attribute this GNMR to the quantum geometric enhancement of diffusion, a novel mechanism distinct from chiral anomaly \cite{Burkov_PRL_2013} or disorder-driven models \cite{Mirlin2001}, in which {\it the quantum metric promotes velocity fluctuations between the spin-split  bands, enhancing diffusion and reducing the resistance}. 
The strength of the effect, $-\beta_{\boldsymbol{g}}(\Delta\epsilon)^2$, is determined by the quantum metric, $\boldsymbol{g}$, encoded in the parameter $\beta_{\boldsymbol{g}}$, and by the energy splitting, $(\Delta\epsilon)^2\sim (\boldsymbol{\cal E}\times{\bf B})^2$, produced by 
$H_{DZ}=-\gamma(\boldsymbol{\cal E}\times{\bf B})\cdot\boldsymbol{\sigma}$, that we term Drift-Zeeman coupling, a novel magnetoelectric-spin interaction that locks the spin of carriers crossing a region of $\boldsymbol{\cal E}\perp{\bf B}$ fields to its drift-momentum, with $\gamma$ determined by the anisotropic spin-orbit interaction. 
Our work therefore establishes a direct manifestation of the non-Markovian memory principle in Te through a previously unexplored quantum mechanical effect: {\it the memory of the wavefunction's quantum geometric structure across the Brillouin zone}. 


\begin{figure}[h!]
\centering
\fbox{\includegraphics[trim={50pt 0pt 50pt 0pt}, clip,angle=-90,width=1\textwidth]{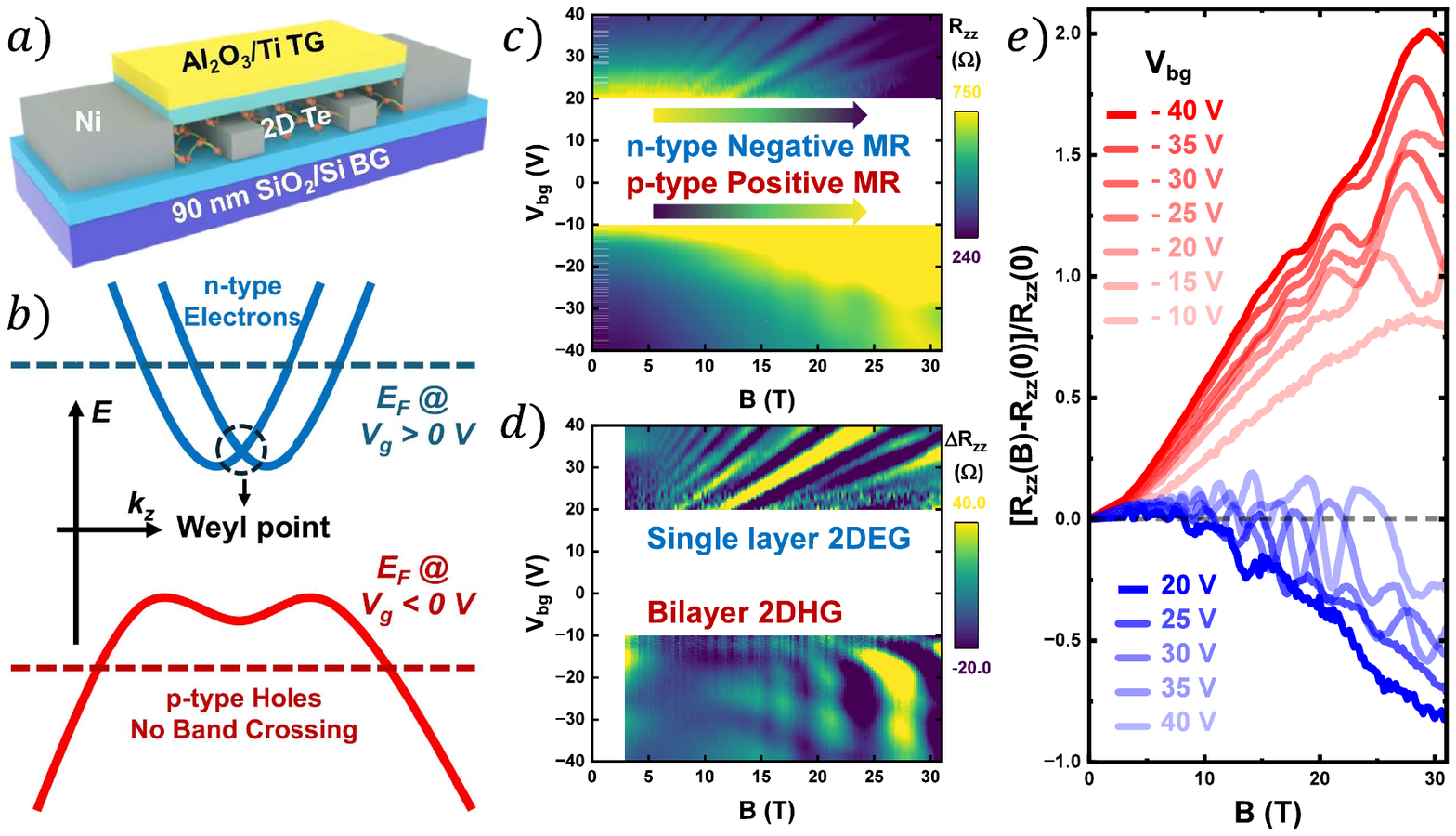}}
\caption{\textbf{Device schematics and gate-tunable quantum transport in $p$- and $n$-type Te.} 
{\bf a)} Schematic illustration of a dual-gated Hall-bar device based on a Te flake for transport measurements. 
{\bf b)} Band structure of Te. At zero gate bias, the Te is nearly intrinsic. Applying a positive gate voltage populates the conduction band, where the spin-split subbands cross at $\mathrm{H}$ point to form a Weyl node. In contrast, a negative gate voltage accesses the valence band without band crossing, enabling comparison between distinct quantum transport regimes governed by the quantum geometry.
{\bf c)} Color map of magnetoresistance ($R_{zz}$) as a function of back-gate voltage and magnetic field, showing negative MR for electron conduction and positive MR for hole conduction within the same device. 
{\bf d)} Background-subtracted $\Delta R_{zz}$ revealing clear SdH oscillations in both $n$-type and $p$-type regimes. While for $n$-type the SdH oscillations are featured by a single characteristic frequency (single-layer 2DEG) for the $p$-type two characteristic frequencies are observed in the SdH oscillations (bilayer 2DHG).
{\bf e)} Overlay of normalized magnetoresistance curves for $p$-type (red) and $n$-type (blue) conduction, highlighting their contrasting field responses.
}
\label{Fig-Pannel-01}
\end{figure}

\section{Experimental Results}
\label{sec:results}

\subsection{Gate-Tunable Quantum transport in Tellurene}
\label{sec:single-double}

We experimentally investigated the magnetotransport properties of Te using Hall-bar devices fabricated on a silicon dioxide substrate, with a highly doped silicon back gate and an $Al_{2}O_{3}$ top gate, as shown in Figure \ref{Fig-Pannel-01}a. Owing to the narrow bandgap of Te \cite{qiu2022resurrection}, both carrier type and carrier density can be continuously tuned within a single device through electrostatic gating. 
Figure \ref{Fig-Pannel-01}b presents the band structure of Te, where the conduction band splits and crosses at the $\mathrm{H}$ point to form a Weyl node, imparting nontrivial quantum geometry to the charge carriers. In contrast, the valence band shows no such crossing. 
This distinct band topology provides an exceptional platform for exploring quantum-geometry-driven transport phenomena in Te. 
Figure \ref{Fig-Pannel-01}c presents a color map of the longitudinal magnetoresistance ($R_{zz}$) as a function of gate voltage (carrier density) and magnetic field. Distinct magnetoresistive behaviors are observed for electrons and holes: the $n$-type regime exhibits NMR, whereas the $p$-type regime shows positive magnetoresistance (PMR).
After subtracting the background resistance, Shubnikov–de Haas (SdH) oscillations emerge in both conduction and valence bands, as shown in Figure \ref{Fig-Pannel-01}d. 
The corresponding Landau-fan diagram reveals two distinct oscillation frequencies in the $p$-type regime, originating from the dual-surface (top and bottom) accumulation layers inherent to Te and consistent with bilayer transport behavior (bilayer 2DHG) \cite{niu2021bilayer}. 
In contrast, the $n$-type regime displays a single oscillation series, consistent with depletion of the top surface and single-layer transport behavior (single-layer 2DEG).
Figure \ref{Fig-Pannel-01}e shows the normalized magnetoresistance measured across different carrier types and densities, revealing a clear transition from PMR to NMR  at a cryogenic temperature of \SI{350}{\milli\kelvin}. 
This gate-tunable NMR persists over a wide range of perpendicular magnetic fields, up to \SI{35}{\tesla}. As the carrier density of the 2DEG increases, shifting the chemical potential away from the Weyl node near the conduction band edge, the magnitude of the NMR gradually diminishes. In contrast, only conventional PMR is observed in the 2DHG under  negative back-gate bias (Figure~\ref{Fig-Pannel-01}f), consistent with the expected 
two surfaces hole accumulation \cite{niu2021bilayer}. 

\begin{figure}[h!]
\centering
\fbox{\includegraphics[trim={60pt 90pt 60pt 90pt}, clip,angle=-90,width=1\textwidth]{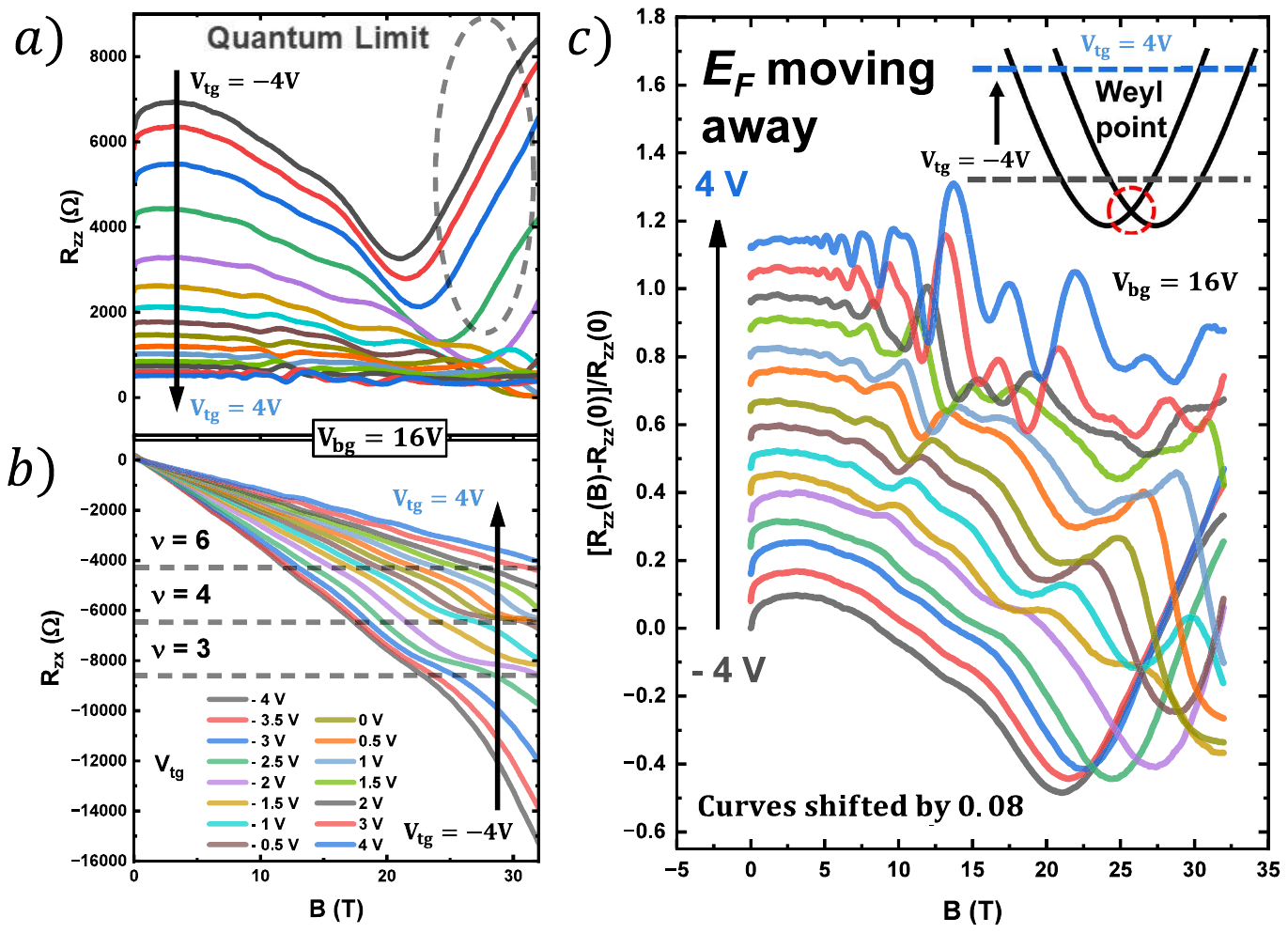}}
\caption{\textbf{Carrier-density dependence of negative magnetoresistance and the quantum Hall effect.}
{\bf a)} Longitudinal magnetoresistance ($R_{zz}$) as a function of carrier density, showing a pronounced negative magnetoresistance (NMR) that persists until the system reaches the quantum limit at the lowest Landau level ($n = 0$).
{\bf b)} Transverse Hall resistance ($R_{zx}$) versus carrier density, revealing well-developed quantum Hall plateaus at filling factors $\nu = 3$, $4$, and $6$.
{\bf c)} Normalized magnetoresistance in the Te conduction band as a function of carrier density. The NMR gradually weakens with increasing gate voltage as the Fermi level shifts away from the Weyl node. $Inset$: Schematic illustration of the carrier-density-dependent Fermi level movement relative to the Weyl node, as the topgate voltage is swept from $-4V$ (black curves in all plots) to $4V$ (dark blue curves in all plots) in the conduction band. 
}
\label{Fig-Pannel-02}
\end{figure}

Figures~\ref{Fig-Pannel-02}a and ~\ref{Fig-Pannel-02}b present the carrier density dependence of the longitudinal and transverse magnetoresistance in the Te conduction band, respectively. Low carrier densities were accessed by fixing the back-gate voltage at $V_{bg}=16$ V, ensuring good contact properties, while tuning the top-gate voltage to modulate the carrier density between $2\times10^{12}$cm$^{-2}$ to $1\times10^{13}$cm$^{-2}$ \cite{QHEWeylTellurene}.
Well developed quantum Hall plateaus of filling factor $\nu = 3$, $4$, and $6$ 
are observed. NMR emerges before the system reaches the quantum limit (Landau level $n = 0$), occurring at high magnetic fields and low carrier densities. 
By continuously increasing the applied magnetic field, the resistance initially decreases, reaches a minimum at a critical crossover field, and subsequently becomes positive. In the quantum limit, the NMR is suppressed as the electrons become localized by cyclotron motion, eliminating interband contributions from the quantum geometric mechanism (as discussed in Section 3).
Figure ~\ref{Fig-Pannel-02}c shows the carrier density dependence of the normalized magnetoresistance, illustrating how the NMR evolves with the Fermi level position relative to the Weyl node in the conduction band.
Our analysis focuses on the magnetic field range between \SI{2}{\tesla} and \SIrange{25}{35}{\tesla}, where the NMR dominates. The weak anti-localization observed below \SI{2}{\tesla} \cite{Chang_Niu_WAL_Tellurene} indicates strong spin-orbit coupling in Te.
Notably, the crossover field between NMR to PMR shifts to higher magnetic fields with increasing carrier density. 
All curves are quantitatively described by a parabolic dependence, $\Delta R_{zz}(B)/R_0\propto- B^2$, where the proportionality  factor varies with the applied voltage, $V$, the orientation of the magnetic field, $\theta$, and the temperature, $T$. 

\subsection{Field Rotation and Temperature Studies}

\begin{figure}[h!]
\centering
\fbox{\includegraphics[trim={60pt 0pt 60pt 0pt}, clip,angle=-90,width=1\textwidth]{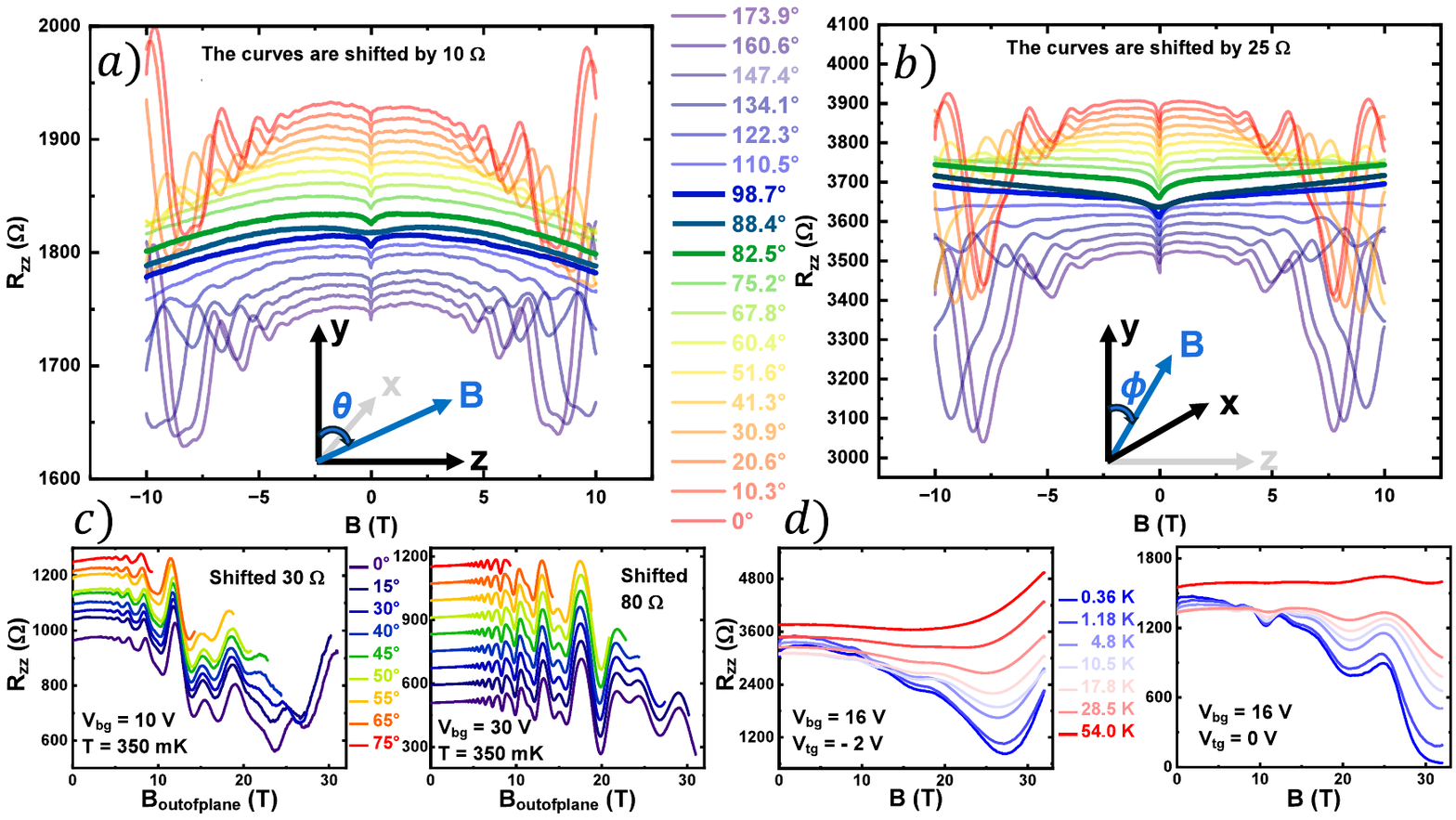}}
\caption{\textbf{Angular, carrier density, and temperature dependence of the NMR.}
{\bf a)} Magnetic field rotation within the $y$-$z$ plane, perpendicular to the device in-plane $x$-axis. 
The angle $\theta$ is measured from the film normal ($\hat{y}$). 
The giant NMR weakens but persists as the field approaches the in-plane $\hat{z}$ direction ($\theta=90^\circ$). {\bf b)} Magnetic field rotation within the $y$-$x$ plane, perpendicular to the device in-plane $z$-axis. 
The angle $\phi$ is measured from the film normal ($\hat{y}$). 
The giant NMR vanishes completely when the field is aligned along the in-plane $\hat{x}$ direction $(\phi=90^\circ)$, which is parallel to the intrinsic polarization field $\boldsymbol{\mathcal{E}}$. 
{\bf c)} Evolution of the SdH oscillation  for tilted magnetic fields ($0\leq\theta \leq 75^\circ$). No significant change in the SdH sequences in low ($V_{\rm bg} = 10$ V) and high ($V_{\rm bg} = 30$ V) carrier densities indicating an ultra small $g$ factor in the Te conduction band.
{\bf d)} Temperature dependence of the NMR at different carrier densities. 
The effect is strongly suppressed with increasing temperature, vanishing entirely at approximately 54 K.
}
\label{Fig-Pannel-03}
\end{figure}

To further elucidate the nature of the observed NMR, we performed a series of field rotation experiments on $n$-type Te devices. 
In Figure~\ref{Fig-Pannel-03}a, the magnetic field is rotated within the $y$-$z$ plane, where the $z$-axis is aligned with the chiral direction of the Te atomic helices and the current flow, and the $y$-axis is perpendicular to the 2D Te flake. 
The data reveal two critical insights: while the SdH oscillations are present only when the field has a component perpendicular to the film, the NMR persists for all field orientations, albeit with varying strength. 
Specifically, the NMR is strongest when the magnetic field is perpendicular to the film, characterized by a tighter parabolic drop in resistance. 
As the field is rotated and becomes aligned with the $z$-axis (parallel to the current), 
the NMR weakens, following a looser parabolic trend. 
As we demonstrate in Sec.~\ref{sec:discussion}, this behavior can be traced back to the anisotropic spin-orbit interaction in Te.

Figure~\ref{Fig-Pannel-03}b presents data from a rotation of the magnetic field within the $y$-$x$ plane. Here, the $x$-axis corresponds to the planar direction perpendicular to the current and, crucially, it is aligned to the large, macroscopic polarization electric field, $\boldsymbol{\cal E}$, which arises from the lone pairs \cite{OurPRL2025}. 
The data show that the NMR completely vanishes when $\vb{B}\parallel\boldsymbol{\cal E}$. 
The combined results from these two rotation scans compellingly 
support that the parabolic magnetoresistance term is proportional to the square of the cross product between $\boldsymbol{\cal E}$ and ${\bf B}$, or $\Delta R_{zz}/R_0=-\beta_g(\boldsymbol{\cal E} \times {\bf B})^2$, a form we derive in Sec.~\ref{sec:discussion}.

Figure~\ref{Fig-Pannel-03}c shows the evolution of the magnetoresistance as a function of the out-of-plane magnetic field for tilted-field configurations in an $n$-type Te device, measured at two representative carrier densities ($V_{\rm bg}$ = 10 V and 30 V).
The data confirm the robustness of the NMR under tilted magnetic fields and highlight its clear dependence on carrier density, with a stronger effect observed at lower carrier densities, corresponding to Fermi levels closer to the Weyl node.
No significant change is found in the angle-dependent SdH oscillation sequences at low ($V_{\rm bg} = 10$ V) and high ($V_{\rm bg} = 30$ V) carrier densities, suggesting an ultra small $g$-factor in the Te conduction band \cite{QHEWeylTellurene} and confirming the absence of ordinary Zeeman splitting under tilted magnetic fields.
Finally, Figure~\ref{Fig-Pannel-03}d further illustrates the temperature dependence of the NMR. 
The effect is strongly suppressed with increasing temperature, completely vanishing around \SI{54}{\kelvin}. This behavior is consistently observed in different carrier densities (and devices), and confirms the quantum geometric origin of the GNMR, as the  geometric interband contribution (dominant at low T) is overshadowed by the thermal intraband one.


\section{Theoretical Discussion}
\label{sec:discussion}

\subsection{Geometric diffusion}

The observation of a gate-tunable, parabolic GNMR in Te exhibiting a distinct $\Delta R_{zz}/R_0 = -\beta_{g}(\boldsymbol{\cal E} \times \mathbf{B})^2$ dependence defies explanation by the mechanisms discussed in Sec. \ref{sec:introduction}. We propose that this phenomenon arises from a quantum geometric enhancement of carrier diffusion, combined with a previously unreported magnetoelectric spin interaction. Our starting point is the Kubo-Greenwood formula \cite{ziman1972principles} for the conductivity tensor, $\sigma_{ij}$, which for an $n$ band system in $d$ dimensions is given by 
\begin{equation}
\sigma_{ij} = e^2 \sum_n \int \frac{d^d\mathbf{k}}{(2\pi)^d} D_{ij}^{(n)}(\mathbf{k}) \left( -\frac{\partial f(\epsilon_n(\mathbf{k}))}{\partial \epsilon_n(\mathbf{k})} \right),
\label{eq:kubo_intro}
\end{equation}
where $e$ is the electric charge, $\epsilon_n(\mathbf{k})$ is the $n-$th band dispersion relation, $f(\epsilon_n(\mathbf{k}))$ is the equilibrium Fermi-Dirac distribution, and $D^{(n)}_{ij}(\mathbf{k})$ is the diffusion tensor, defined as the time integral of the velocity auto-correlation \cite{mcquarrie1976statistical} between Bloch states $| u_n(\mathbf{k}) \rangle$
\begin{equation}
D_{ij}^{(n)}(\mathbf{k}) = \int_0^{\infty} dt \, \langle u_n(\mathbf{k}) | v_i(t) v_j(0) | u_n(\mathbf{k}) \rangle.
\label{non-Markovian}
\end{equation}
In a multi-band system, like the conduction band of Te, the velocity operator comprises a conventional intraband group velocity and an interband component responsible for quantum geometric fluctuations, $v_i(t) = \partial_{k_i} \epsilon(\mathbf{k})/\hbar + \delta v_i(t)$. Evaluating the correlation function $\langle \delta v_i(t) \delta v_j(0) \rangle \sim \langle \delta v_i \delta v_j \rangle e^{-\Gamma_g t}$, where $\tau_g = 1/\Gamma_g$ is a temperature independent quantum geometric relaxation time, yields the central result for the diffusion tensor $D_{ij}(\mu)=\sum_{n=\pm}D_{ij}^{(n)}(\mu)$ at the chemical potential $\mu$
\begin{equation}
D_{ij}(\mu) = \underbrace{D_{ij}^D(\mu)}_{\text{intraband}} + \underbrace{\frac{2\tau_{g}}{\hbar^2} \langle[\Delta \epsilon({\bf k})]^2 g_{ij}({\bf k})\rangle_{FS}}_{\text{interband (geometric)}}.
\label{eq:D_full}
\end{equation}
Here, $D_{ij}^D(\mu)$ is the Drude (intraband) Markovian contribution to the diffusion tensor \cite{ziman1972principles} and $\langle[\Delta \epsilon({\bf k})]^2 g_{ij}({\bf k})\rangle_{FS}$ is the Fermi-surface average of the band-splitting squared times the quantum metric $g_{ij}(\mathbf{k}) = \text{Re}[\langle \partial_{k_i} u | \partial_{k_j} u \rangle - \langle \partial_{k_i} u | u \rangle \langle u | \partial_{k_j} u \rangle]$, that quantifies the overlap between Bloch states in ${\bf k}-$space \cite{QGinCMLiu2024,Kaplan2024}. This geometric contribution that arises from the time-integrated velocity auto-correlation in Eq. (\ref{non-Markovian}) is a $\mathbf{k}$-space  effect that is fundamentally non-Markovian, a memory of the carrier's past history encoded in its quantum geometric structure. It influences the carrier's trajectory by enhancing diffusion and reducing the resistance, while the field dependence of the band splitting will ultimately describe the experimental GNMR. The novel mechanism reported here is the transport analog of the quantum geometric enhancement of the superfluid weight in conventional \cite{Wei_Chen_2005} and flat-band superconductors \cite{Peotta2015,Liang2020}.

%
\begin{figure}[h!]
\centering
\fbox{\includegraphics[trim={50pt 0pt 50pt 0pt}, clip,angle=-90,width=1\textwidth]{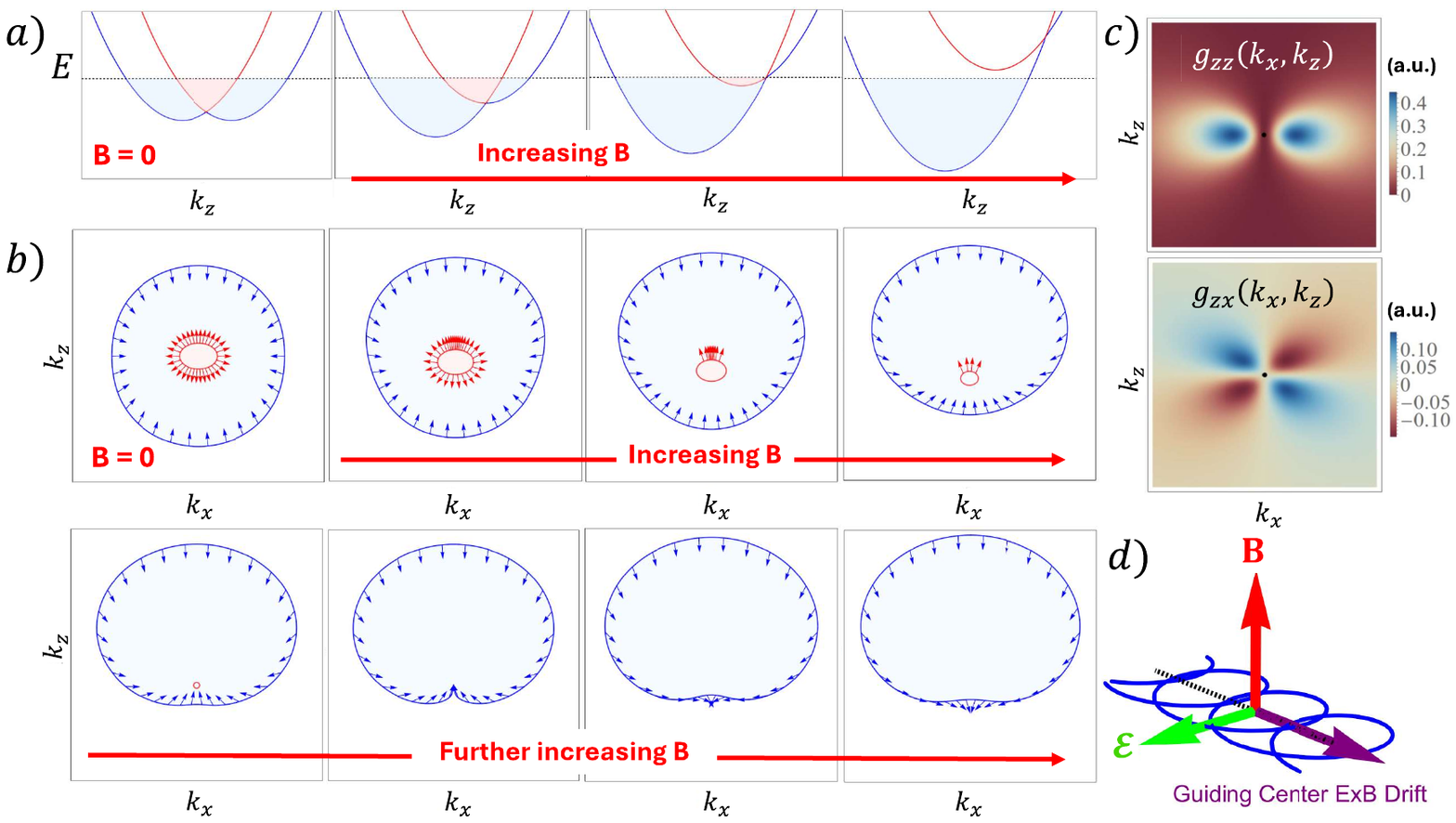}}
\caption{\textbf{Quantum geometry and spin-drift locking in Te.}
\textbf{(a)} Conduction bands $\epsilon_\pm(k_z)$ of Te at a given chemical potential, featuring a Weyl node located at $k_z=0$, which is the origin of its unique electronic properties, and its evolution with magnetic field, $\boldsymbol{B}\parallel\hat{\boldsymbol{y}}$.
\textbf{(b)} Evolution of the radial configuration of the spin texture (blue and red arrows for the outer and inner Fermi surfaces) as the magnetic field is increased.  
\textbf{(c)} Quantum metric components, $g_{zz}(\boldsymbol{k})$ and $g_{zx}(\boldsymbol{k})$.
\textbf{(d)} The locking of the spin to the guiding-center drift upon crossed electric $\boldsymbol{\cal E}$ and magnetic $\boldsymbol{B}$ fields. }
\label{Fig-Pannel-04}
\end{figure}
%

\subsection{The drift-Zeeman interaction}

The geometrically enhanced diffusion given by Eq.~(\ref{eq:D_full}) is activated by a magnetic-field-dependent energy splitting, $\Delta \epsilon(\mathbf{k})$, between the two non-degenerate bands (Fig.~\ref{Fig-Pannel-04}a). The ultra small $g-$factor in the conduction band of Te \cite{QHEWeylTellurene} indicates the absence of the ordinary Zeeman effect, and a different mechanism for band splitting must be operational. 
We identify its source as a novel magnetoelectric spin interaction, which emerges from the interplay between Te's intrinsic polarization electric field $\boldsymbol{\cal E}\approx({\cal E}_x,0,0)$ (from lone pairs) \cite{OurPRL2025} and an applied magnetic field $\mathbf{B}$. The unperturbed Weyl Hamiltonian for the Te conduction band near the $\mathrm{H}$ point is
\begin{equation}
H_0 = \frac{\hbar^2 k_\perp^2}{2m_\perp^*} + \frac{\hbar^2 k_z^2}{2m_z^*} + \lambda_\perp (k_x \sigma_x + k_y \sigma_y) + \lambda_z k_z \sigma_z,
\label{eq:H_full}
\end{equation}
where $k_\perp=\sqrt{k_x^2+k_y^2}$, $m_\perp^*$ and $m_z^*$ are the effective masses perpendicular and parallel to the helical direction, respectively, and $\lambda_\perp$ and $\lambda_z$ are the anisotropic spin-orbit couplings. The electric field due to lone pairs $\boldsymbol{\cal E}$ couples through the dipole interaction $H_E = -e \boldsymbol{\cal E} \cdot \mathbf{r}$ \cite{sakurai2020modern}. The magnetic field $\mathbf{B}$, instead, is introduced via the Peierls substitution, $\mathbf{k} \rightarrow \mathbf{k} - (e/\hbar)\mathbf{A}$  \cite{landau1977quantum}, using the symmetric gauge $\mathbf{A} = \frac{1}{2} (\mathbf{B} \times \mathbf{r})$, which yields a term given by $H_B = -\frac{e \lambda}{2\hbar} (\mathbf{B} \times \mathbf{r}) \cdot \boldsymbol{\sigma}$. Both $H_E$ and $H_B$ depend on the position operator $\mathbf{r}$, whose matrix elements are evaluated using the commutation relation $[ \mathbf{r}, \lambda{\bf k}\cdot\boldsymbol{\sigma} ] = i \lambda  \boldsymbol{\sigma}$. Treating $H_E$ and $H_B$ as perturbations to $H_0$, we use second-order perturbation theory \cite{sakurai2020modern} to calculate the cross term $\langle 0|H_E|n\rangle\langle n|H_B|0\rangle+\text{H.c.}$, which leads to the Drift-Zeeman Hamiltonian (see Supplementary Information for details)
\begin{equation}
H_{DZ} = -\gamma \cdot (\boldsymbol{\cal E} \times \mathbf{B}) \cdot \boldsymbol{\sigma}.
\label{eq:Drift_Zeeman_interaction}
\end{equation}
The anisotropy of this novel magnetoelectric spin-interaction is captured by the vector $\gamma=(\gamma_\perp,\gamma_\perp,\gamma_z)$, where $\gamma_z = 2e^2\lambda_z^3/(\hbar\Delta^3)$ and $\gamma_\perp = 2e^2\lambda_\perp^3/(\hbar\Delta^3)$, with $\gamma_z>\gamma_\perp$, while $\Delta$ is a typical energy bandwidth that characterizes the Te band structure. 

\subsection{The negative magnetoresistance}

The Drift-Zeeman interaction (\ref{eq:Drift_Zeeman_interaction}) has a profound impact on the electronic structure of Te. For $\mathbf{B} \parallel \hat{y}$ and $\boldsymbol{\cal E} \parallel \hat{x}$ \cite{OurPRL2025}, it shifts the Weyl node along the $k_z$-axis (Fig.~\ref{Fig-Pannel-04}a), modifying the Fermi surface topology and spin texture, as shown in Fig.~\ref{Fig-Pannel-04}b \cite{focassio2024magnetic}. The states from the inner Fermi surface are transferred to the outer Fermi surface, while the overall spin texture reorients itself to align with the effective $\boldsymbol{\cal E} \times \mathbf{B}$ field. As a result, the quantum geometric contribution to the diffusive transport (\ref{eq:D_full}) is given by 
\begin{eqnarray}
    \langle [\Delta \epsilon({\bf k})]^2 g_{zz}({\bf k})\rangle_{FS}&=&-8\underbrace{\gamma_z\langle k_z g_{zz}({\bf k})\rangle_{FS}\,\hat{z}\cdot(\boldsymbol{\cal E}\times{\bf B})}_{\mbox{anti-symmetric}}\nonumber\\
    &+&\underbrace{4\langle(\lambda_\perp^2 k_x^2 + \lambda_z^2 k_z^2)g_{zz}({\bf k})\rangle_{FS}+4\langle g_{zz}({\bf k})\,[\gamma\cdot(\boldsymbol{\cal E}\times{\bf B})]^2\rangle_{FS}}_{\mbox{symmetric}}.
    \label{eq:symm-and-anti-symm-contributions}
\end{eqnarray}
If we neglect the ${\bf B}$ dependence of the quantum metric, then $g_{ij}(-{\bf k})=g_{ij}({\bf k})$ and the anti-symmetric contribution becomes identically zero, $\langle k_z g_{zz}({\bf k})\rangle_{FS}= 0$ because $k_z$ is odd but $g_{zz}(k_x,k_z)$ is even, see Fig.~\ref{Fig-Pannel-04}c (top). In this case, the symmetric contribution produces the parabolic GNMR, $\Delta R_{zz}(B)/R_0=-\beta_g(\boldsymbol{\cal E}\times{\bf B})^2$, observed experimentally, with a curvature given by $\beta_g\propto \langle g_{zz}({\bf k})\rangle_{FS}$. Furthermore, no Hall contribution from $\langle g_{zx}({\bf k})\rangle_{FS}\rightarrow 0$ is expected, due to the symmetry of $g_{zx}({\bf k})$, see Fig.~\ref{Fig-Pannel-04}c (bottom).

Finally, in the quantum limit the GNMR is expected to be suppressed as the interband contributions saturate. In this regime the crystal momentum $\mathbf{k}$ is no longer a good quantum number, being substituted by the quantized Landau levels $n$. The quantum metric, which previously scaled as $g_{\mu\nu}({\bf k})\sim 1/|{\bf k}|^2$, now becomes governed by the magnetic length, $\ell_{B}=\sqrt{\hbar/eB}$, and scales as $g_{\mu\nu}(B) \propto \ell_B^2 \propto 1/B$. At the $n-$th Landau level, the band-splitting scales as $\Delta E_n(B)\sim\lambda\sqrt{n}/\ell_B\sim \sqrt{B}$, and therefore the quantum geometric contribution diffusion, $D_{ij} \propto g_{\mu\nu} (\Delta E_n)^2\sim n$, saturates, suppressing the GNMR. In the extreme quantum limit, when only the non-degenerate $n=0$ Landau level is filled, $\Delta E_0=0$ and there is no geometric contribution to diffusion.

\subsection{The spin-drift locking}

The interaction term we put forward (\ref{eq:Drift_Zeeman_interaction}) has a very elegant and intuitive physical interpretation. 
For a 2DEG in the $x$-$z$ plane and perpendicular magnetic field, ${\bf B}\parallel\hat{y}$, the energy of an electron is independent of the guiding center position of the cyclotron orbit, $(X_0, Z_0)$, resulting in a massive degeneracy of $eB/h$ states per unit area \cite{Levitov2008}. When the lone-pair electric field $\boldsymbol{\cal E} = {\cal E} \, \hat{x}$ is added to the problem, a potential energy $-e{\cal E} X_0$ breaks such degeneracy, localizing the $X_0$ coordinate and promoting the drift of the guiding center orbits along the $Z_0$ direction, with a drift momentum ${\bf k}_d\propto\boldsymbol{\cal E}\times{\bf B}$, see Fig.~\ref{Fig-Pannel-04}d. 
For a material with strong spin-orbit interaction, $H_{s.o.} = \lambda \, \mathbf{k} \cdot \boldsymbol{\sigma}$, such as in Te, the drifting Weyl fermions feature a well-defined spin orientation $\langle \boldsymbol{\sigma}\rangle \parallel \langle \mathbf{k}_d \rangle \propto \boldsymbol{\cal E} \times \mathbf{B}$. 
This is the spin-drift-momentum locking described by $H_{DZ}$
(\ref{eq:Drift_Zeeman_interaction}).

\section{Comparison to experiments}

\begin{figure}[ht!]
\centering
\fbox{\includegraphics[trim={70pt 0pt 60pt 0pt}, clip,angle=-90,width=\linewidth]{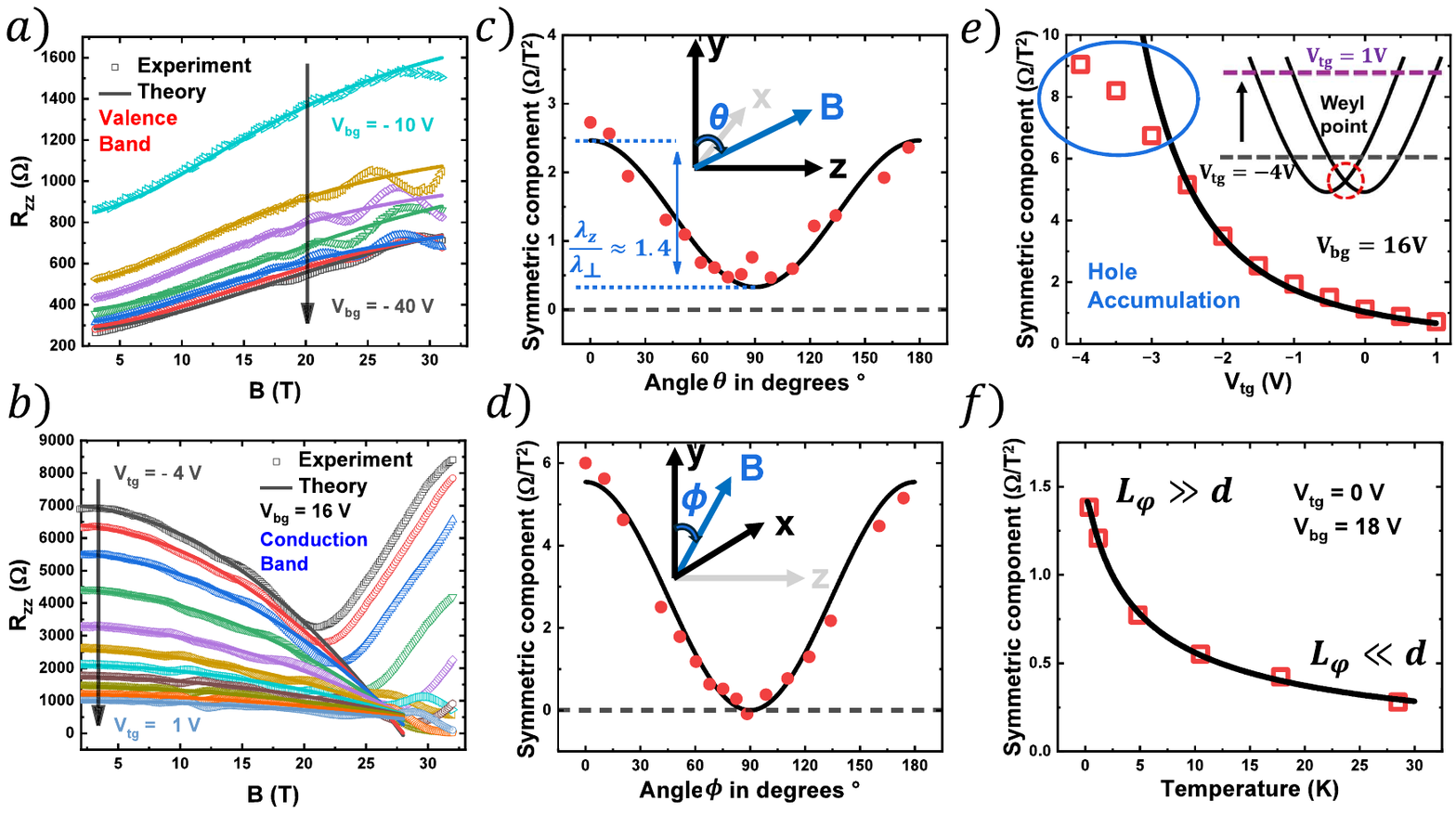}}
\caption{\textbf{Quantitative analysis of the GNMR in Te.}
\textbf{(a)} Fit of the PMR for $p$-type carriers in the valence band, from $V_{bg}=-10$V to $V_{bg}=-40$V and $V_{tg}=0$V, following the expected behavior for two hole accumulation layers.
\textbf{(b)} Fit of the NMR for $n$-type carriers in the conduction band, from $V_{tg}=-4$V to $V_{tg}=+4$V and $V_{bg}=+16$V, following the parabolic enhanced diffusion from quantum geometry.
\textbf{(c)} Angular evolution of the symmetric component $C(\theta)$ of the GNMR for scan rotations of the magnetic field, ${\bf B}$, confined to the $y$-$z$ plane. The ratio $C(0)/C(\pi/2)=\lambda_z^6/\lambda_\perp^6\approx 7.5$ is a direct measure of the anisotropy in the spin-orbit couplings $\lambda_z/\lambda_\perp\approx 1.4$. 
\textbf{(d)} Angular evolution of $C(\phi)$ of the GNMR for scan rotations of the magnetic field, ${\bf B}$, confined to the $y$-$x$ plane.  
\textbf{(e)} Evolution of $C(\theta=0)$ of the GNMR as a function of the top-gate voltage, $V_{tg}=-4$V to $V_{tg}=+1$V and $V_{bg}=+16$V, decreasing as the Fermi level is tuned away from the Weyl node (inset), consistent with a quantum geometric origin. For $-4$V$<V_{tg}<-3$V, at the bottom of the conduction band and very close to the Weyl node, hole accumulation occurs spoiling the agreement between theory and experiment.
\textbf{(f)} Evolution of $C(\theta=0)$ of the GNMR as a function of temperature, $T=0.35$K to $T=54$K, at $V_{bg}=+18$V and $V_{tg}=0$V. The data (circles) are well described by $C(T) = C_0/(1 -\beta_{wl}\ln{(T/T_0)}+ \alpha_{ee}\sqrt{T})^2$ (solid line), describing quantum corrections to conductivity through the dimensional crossover 2D $\rightarrow$ 3D, for $L_\varphi\gg d$ and $L_\varphi\ll d$.}
\label{Fig-Pannel-05}
\end{figure}

\noindent
{\bf Positive magnetoresistance} $-$ We used negative back-gate voltages, $V_{bg}<0$, to tune the Fermi level in the valence band of Te. The data for the observed longitudinal PMR shown in Fig. \ref{Fig-Pannel-05}a for $-40$ V $< V_{bg}< -10$ V, are well described by the formula $R_{zz}(B) = R_0 \left(1 + \frac{\sigma_1 \sigma_2 (\mu_1 - \mu_2)^2 B^2}{(\sigma_1 + \sigma_2)^2 + (\sigma_1 \mu_2 + \sigma_2 \mu_1)^2 B^2}\right)$ \cite{niu2021bilayer}, which arises from the coexistence of two distinct $p$-type channels of coefficients $\mu_1,\sigma_1$ and $\mu_2,\sigma_2$, with $\mu_1\neq\mu_2$, associated with the two hole accumulation layers at the surfaces of the device \cite{niu2021bilayer}.

\noindent
{\bf Negative magnetoresistance} $-$ For the conduction band, we used a fixed positive back-gate voltage, $V_{bg}=+16$V, and variable top-gate voltage, $-4$V $<V_{tg}<+4$V, to tune the concentration of $n$-type carriers. 
The longitudinal resistance $R_{zz}(B)$ was measured as a function of $B$, $\theta$, $V$, and $T$. 
The data shown in Fig. \ref{Fig-Pannel-05}b exhibit a clear NMR whose non-oscillatory part of can be fitted using the quantum geometric expression given in Eq. (\ref{eq:symm-and-anti-symm-contributions}), recast as  $R_{zz}(B)=R_0+F\cdot B-C \cdot B^2$, with $F$ and $C$ representing, respectively, the anti-symmetric (in units of $\Omega/T$) and symmetric (in units of $\Omega/T^2$) components of the magnetoresistance. 
Equation~\eqref{eq:symm-and-anti-symm-contributions} predicts that $F=0$,
so any observed anti-symmetric component is entirely due to experimental procedures, such as, for example, sample alignment. 
As such, in what follows we describe the evolution of the symmetric component (see Supplementary Information for details) 
\begin{equation}
C=\frac{(L/W)\; e^2N(E_F) \; (8\tau_g/\hbar^{2}) \; \langle g_{zz}({\bf k}) \rangle_{FS}\;\mathcal{E}_x^{2}}{ \left( \sigma_{zz}^D + \sigma_{zz}^{g}\right)^{2} }(\gamma_z^{2} \cos^2\alpha_{(\mathbf{B},\hat{y})}+\gamma_\perp^2\cos^2\alpha_{(\mathbf{B},\hat{z})} ),
\label{eq:symm-component-C}
\end{equation}
as a function of $\alpha_{(\mathbf{B},\hat{y})}$ and $\alpha_{(\mathbf{B},\hat{z})}$, $V$, and $T$, where $L$ and $W$ are the dimensions of the Te film,  $N(E_F)$ is the density of states at the Fermi level, and $\sigma_{zz}^{g}$ is the ${\bf B}=0$ quantum geometric contribution to conductivity, $\sigma_{zz}^{g}=4e^2N(E_F)\langle(\lambda_\perp^2 k_x^2 + \lambda_z^2 k_z^2)g_{zz}({\bf k})\rangle_{FS}$. 

\noindent
{\bf Angular dependence} $-$ For the $y$-$z$ rotations, $\alpha_{(\mathbf{B},\hat{y})}=\theta$ and $\alpha_{(\mathbf{B},\hat{z})}=\pi/2-\theta$, the symmetric component $C(\theta)$ in Fig. \ref{Fig-Pannel-05}c follows the quantum geometric prediction (\ref{eq:symm-component-C}), with $C(\theta)=C_z\cos^2\theta+C_\perp\sin^2\theta$. 
The ratio $C_z/C_\perp\equiv\gamma_z^2/\gamma_\perp^2=\lambda_z^6/\lambda_\perp^6\approx 7.5$ is a direct measure of the anisotropic spin-orbit interaction, $\lambda_z/\lambda_\perp\approx 1.4$, and agrees well with previous experimental results from magnetooptical transitions in the conduction band of Te, $\lambda_z/\lambda_\perp\approx 1.6$ \cite{Magnetooptical}, and also with the fitted band structure from DFT ($\lambda_z/\lambda_\perp\approx 1.8$). For the $y$-$x$ rotations, instead, $\alpha_{(\mathbf{B},\hat{y})}=\phi$ and $\alpha_{(\mathbf{B},\hat{z})}=\pi/2$, so $C(\phi)$ in Fig. \ref{Fig-Pannel-05}d now follows $C(\phi)=C_z\cos^2\phi$, vanishing at $\phi=\pi/2$. 
The angular dependencies described above establish unambiguously that the parabolic GNMR is maximized for $\mathbf{B} \perp \boldsymbol{\cal E}$ and vanishes for $\mathbf{B} \parallel \boldsymbol{\cal E}$, confirming the structure, $\Delta R_{zz}/R_0 \propto -[\gamma\cdot(\boldsymbol{\cal E} \times \mathbf{B})]^2$, predicted by quantum geometry. 

\noindent
{\bf Voltage dependence} $-$ In the limit where the Drude contribution $\sigma_{zz}^D \gg \sigma_{zz}^g$, and using a parabolic band approximation in Eq. (\ref{eq:symm-component-C}) for the Fermi surface average, $\langle g_{zz}({\bf k})\rangle_{FS}$, we obtain the scaling $C(\theta=0) \propto V^{-5/2}$ for the voltage dependence of the symmetric component $C(\theta=0,V_{bg})$ shown in Fig.~\ref{Fig-Pannel-05}e (see Supplementary Information for details). This unique dependence with $V$ arises from the scaling $\langle g_{zz}({\bf k}) \rangle_{\rm FS} \sim 1/\sqrt{n}$, and the fact that $C\propto 1/(\sigma_{zz}^D)^2$, with the Drude conductivity scaling as $\sigma_{zz}^D \sim n$, and recalling that in 2D (where $N(E_F)$ is constant) we have $n\propto V$. The rapid decay of $C(\theta=0)$ with increasing voltage (density) shown in Fig. \ref{Fig-Pannel-05}e demonstrates that quantum geometric effects are most pronounced in low-carrier-density regimes, due to the proximity to a Weyl node. However, for $-4$V$<V_{tg}<-3$V, at the bottom of the conduction band and close to the Weyl node, hole accumulation at the top surface occurs compromising the direct comparison between theory and experiment.  

\noindent
{\bf Temperature dependence} $-$ The temperature dependence of $C$ in Eq. (\ref{eq:symm-component-C}) is characteristic of a disordered metal undergoing dimensional crossover, 2D $\rightarrow$ 3D, when the phase coherence length, $L_\varphi\sim T^{-1/2}$, becomes smaller than the thickness of the film, $d$. The model $C(T) = C_0/(1 -\beta_{WL}\ln{(T/T_0)}+ \alpha_{ee}\sqrt{T})^2$ results from the symmetric component, $C$, being inversely proportional to the square of the Drude conductivity, $C\propto 1/(\sigma_{zz}^D(T))^2$, with $\delta\sigma_{2D}(T)=-\sigma_{WL}\ln{(T/T_0)}$ describing the quantum corrections to conductivity due to WL in 2D, valid when $L_\varphi\gg d$, and $\delta\sigma_{3D}(T)=+\sigma_{ee}\sqrt{T}$ describing the quantum corrections to conductivity due to electron-electron interactions in 3D (when exchange contributions dominate over the Hartree contributions \cite{oliveira2021pressure}), valid when $L_\varphi\ll d$ \cite{Altshuler1980}. Although at $T<1$K and $V_{bg}=+30$V we find $L_\varphi\sim 500$nm \cite{Chang_Niu_WAL_Tellurene}, at lower $V_{bg}=+18$V and higher $T>20$K the quantum interference phenomena in our $d=10-20$ nm Te films crosses over from the 2D ($-\ln{T}$) behavior to the 3D ($+\sqrt{T}$) behavior, as can be seen in Fig. \ref{Fig-Pannel-05}f. 

\section{Conclusion}\label{sec13}

The remarkable phenomenon of the quantum geometric GNMR reported here signals a new paradigm for quantum transport in 2D materials. Our field rotation and gate voltage dependence studies provide definitive evidence that this effect is not reducible to conventional WL or semiclassical dynamics. Instead, our work establishes a direct manifestation of the memory principle through a previously unexplored quantum mechanical effect: the memory of the wavefunction’s quantum geometric structure across the Brillouin zone. This provides a persistent ``memory" that dictates the macroscopic transport properties. Furthermore, the excellent agreement between our experimental data and our theoretical model of a novel, quantum-geometric-induced Drift-Zeeman interaction opens a new chapter in the study of electron-field interactions in topological quantum matter.

Unlike giant magnetoresistance (GMR) sensors, which saturate at moderate fields, Te’s parabolic GNMR could provide non-saturating response over a wide field range. If the GNMR can be coupled with spin-orbit torques (e.g., in Te/ferromagnet heterostructures), it could enable energy-efficient MRAM with field-programmable resistance states. The absence of fast saturation in our devices suggests multi-bit storage potential. The quantum metric-driven GNMR could be harnessed to couple superconducting qubits to high-field control lines, leveraging the geometric contribution to engineer noise-resilient readout schemes.

\section{Methods}
\label{sec:methods}

\subsection{Sample preparation and electrical transport measurements}

Two-dimensional Te were synthesized via a hydrothermal growth method. A precursor solution of Na$_2$TeO$_3$ and polyvinylpyrrolidone (PVP) in deionized water was prepared, followed by the addition of aqueous ammonia and hydrazine hydrate. The mixture was sealed in a Teflon-lined autoclave and reacted at $180^\circ$C for 12–30 hours before being cooled to room temperature. The resulting 2D Te flakes were transferred onto SiO$_2$/Si substrates (90 nm thermal oxide) using the Langmuir-Blodgett technique to ensure uniform, clean films. Flake thicknesses ranged from a few atomic layers to ~20 nm.

Hall bar devices were fabricated using standard electron beam lithography. Metal contacts (Ni/Au) were deposited via electron beam evaporation. To enable top-gating, a 20 nm Al$_2$O$_3$ dielectric layer was deposited by atomic layer deposition (ALD) at $200^\circ$C. The devices exhibited high electronic quality, with tunable carrier densities ($n \sim 1\times10^{12}$ to $1.2\times10^{13}$ cm$^{-2}$) and high mobilities ($\mu \sim 6000$ cm$^2$/V$\cdot$s for electrons). Well-defined Shubnikov-de Haas oscillations confirmed low defect concentrations.

Transport measurements were performed in high-field magnet systems (cell 9 and cell 12) at the National High Magnetic Field Laboratory (NHMFL), Florida, USA, with temperatures ranging from 350 mK to 54 K and magnetic fields up to 42 T. Longitudinal ($R_{zz}$) and Hall ($R_{zx}$) resistances were measured using phase-sensitive lock-in techniques (SR830/SR860 amplifiers) with low-frequency AC excitation (7–87 Hz) to minimize capacitive coupling.

\subsection{Electronic structure and spin texture}

\begin{figure}[h!]
\centering
\fbox{\includegraphics[trim={20pt 5pt 20pt 10pt}, clip,width=1\textwidth]{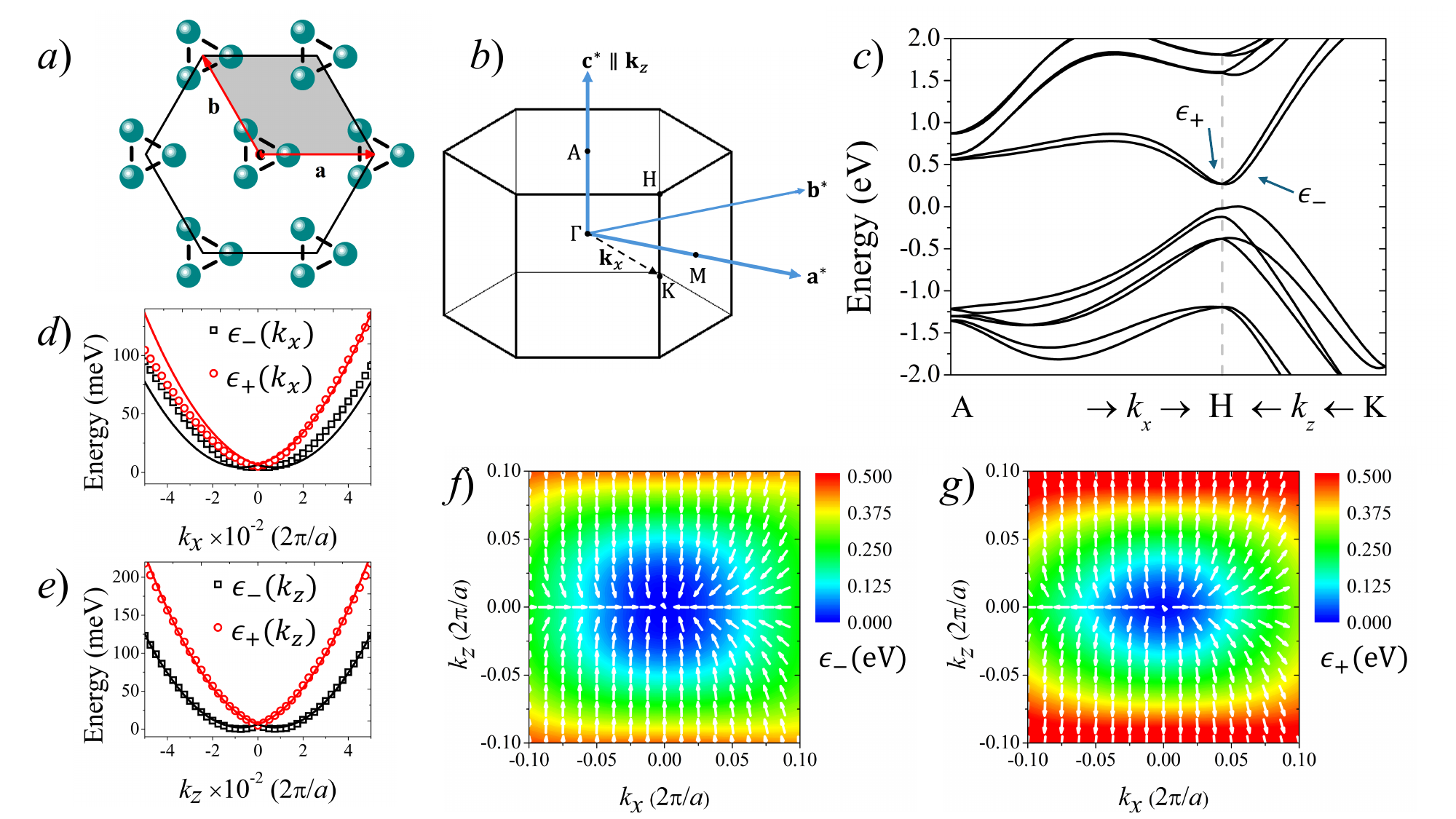}}
\caption{(a) Crystal structure of Te, showing the helical atomic chains running along the $z$-axis, parallel to the $\mathbf{c}$ vector. (b) Hexagonal Brillouin zone of Te with labeled high-symmetry points. (c) Electronic band structure of bulk Te obtained from DFT calculations, highlighting the Weyl node at the $\mathrm{H}$ point arising from the crossing of the first two conduction bands, $\epsilon_{\pm}$, near their minimum. (d) and (e) Fits of the DFT dispersion relations $\epsilon_{\pm}(k_{x},k_{z})$ to the eigenvalues of the unperturbed model Hamiltonian. The excellent agreement along $k_{z}$ in (e) enables the extraction of the ratio $\lambda_{z}/\lambda_{\perp} \approx 1.8$. In contrast, the poor agreement along $k_{x}$ in (d) reflects the influence of trigonal warping, which is not included in the present model. (f) and (g) Two-dimensional colormap dispersions for the lower band $\epsilon_{-}(k_{x},k_{z})$ and the upper band $\epsilon_{+}(k_{x},k_{z})$, together with their corresponding radial spin textures, pointing inward for $\epsilon_{-}$ and outward for $\epsilon_{+}$.
}
\label{Fig-Pannel-06}
\end{figure}

The crystal structure of Te consists of helical chains running along the $z$-axis (Fig.~\ref{Fig-Pannel-06}a), with each atom featuring covalent bonds to two neighbors and a lone pair of electrons extending perpendicular to the chain ($x$-axis) \cite{OurPRL2025}. This structure breaks inversion symmetry and generates a significant intrinsic electric polarization field, $\boldsymbol{\cal E}$, oriented along the $x$-axis.
The Brillouin zone is a hexagonal prism (Fig.~\ref{Fig-Pannel-06}b), and the band structure features a Weyl node near conduction band minimum at the $\mathrm{H}$ point (Fig.~\ref{Fig-Pannel-06}c). For $n$-type doping, the Fermi level resides in the conduction band, which exhibits linear dispersion and strong spin-orbit coupling, characteristic of a Weyl semiconductors. This leads to substantial Berry curvature and a non-trivial quantum metric. The conduction band is composed by two branches $\varepsilon_\pm(k_x,k_z)$ shown in Fig.~\ref{Fig-Pannel-06}d and Fig.~\ref{Fig-Pannel-06}e for two different directions in the Brillouin zone, and these branches are, in turn, associated to opposite total spin textures, shown in Fig.~\ref{Fig-Pannel-06}f and Fig.~\ref{Fig-Pannel-06}g.

Density-functional theory (DFT) calculations were performed within the generalized gradient approximation for the exchange correlation functional using the \textsc{Quantum Espresso} package \cite{Giannozzi_2009}, with fully relativistic Perdew-Burke-Ernzerhof pseudopotentials to account for spin-orbit coupling. Optimal structural and convergence parameters were taken from Ref. \cite{oliveira2021pressure}, adopting lattice constants $a = 4.767$~\AA\ and $c/a = 1.2447$ for the hexagonal cell. Te atoms form helical chains along $z$, with three atoms per unit cell separated by $u = 0.245a$. For the self-consistent calculations, a dense $24 \times 24 \times 20$ Monkhorst Pack $k$-point mesh and a plane-wave cutoff energy of $50$Ry were employed. For the band-structure and spin-texture calculations, a fine mesh of $1 \times 90 \times 80$ $k$-points around the $\mathrm{H}$ point was used to generate the two-dimensional electronic and spin maps, as well as the dispersions along $k_{x}$ and $k_{z}$.

To extract the ratio $\lambda_{z}/\lambda_{\perp}$, we fit the DFT conduction bands to the analytical eigenvalues of Eq. \eqref{eq:H_full}, given by
\begin{equation}
\epsilon_{\pm}(\mathbf{k})
= \frac{\hbar^{2}k_{\perp}^{2}}{2m^{*}_{\perp}} 
+ \frac{\hbar^{2}k{z}^{2}}{2m^{*}_{z}} 
\pm \sqrt{\lambda_{\perp}^{2} k_{\perp}^{2} + \lambda_{z}^{2} k_{z}^{2}} ,
\end{equation}
where the wave-vector components are defined with respect to the $\mathrm{H}$ point, and we take $k_{\perp} = \sqrt{k_{x}^{2} + k_{y}^{2}}$ with $k_{y} = 0$ for the fitting procedure.

\backmatter

\bmhead{Supplementary Information}

Supplementary Information can be found here.

\bmhead{Acknowledgements}
V.V. acknowledges financial support of PNRR MUR project PE0000023-NQSTI and PRIN 2022 (Prot. 20228YCYY7). 
A portion of this work was performed at the National High Magnetic Field Laboratory, which is supported by National Science Foundation Cooperative Agreement No. DMR-1644779 and DMR-2128556* and the State of Florida.
This work is supported by the Brazilian funding agencies FAPERJ and CNPq.

\bibliography{GNMR-QG-ArXive-Bibliography}

\end{document}



\title[Article Title]{Supplementary Information to Gate-Tunable Giant Negative Magnetoresistance in Tellurene Driven by Quantum Geometry}


\author*[1]{\fnm{Marcello B.} \sur{Silva Neto}}\email{mbsn@if.ufrj.br}
\equalcont{These co-first-authors contributed equally to this work.}

\author[2,3]{\fnm{Chang} \sur{Niu}}\email{niu43@purdue.edu}
\equalcont{These co-first-authors contributed equally to this work.}

\author[4]{\fnm{Marcus V. O.} \sur{Moutinho}}\email{moutinho@xerem.ufrj.br}

\author[5]{\fnm{Pierpaolo} \sur{Fontana}}\email{pierpaolo.fontana@uab.cat}

\author[6]{\fnm{Claudio} \sur{Iacovelli}}\email{claudio.iacovelli@hotmail.it}

\author[7]{\fnm{Victor} \sur{Velasco}}\email{vvelasco@sissa.it}

\author[1]{\fnm{Caio} \sur{Lewenkopf}}\email{lewenkopf@if.ufrj.br}

\author*[2,3]{\fnm{Peide D.} \sur{Ye}}\email{yep@purdue.edu}

\affil[1]{\orgdiv{Instituto de F\'isica}, \orgname{Universidade Federal do Rio de Janeiro}, \orgaddress{\city{Rio de janeiro}, \postcode{21941-972}, \state{RJ}, \country{Brazil}}}

\affil[2]{\orgdiv{Elmore Family School of Electrical and Computer Engineering}, \orgname{Purdue University}, \orgaddress{\city{West Lafayette}, \postcode{47907}, \state{Indiana}, \country{United States}}}

\affil[3]{\orgdiv{Birck Nanotechnology Center}, \orgname{Purdue University}, \orgaddress{\city{West Lafayette}, \postcode{47907}, \state{Indiana}, \country{United States}}}

\affil[4]{\orgname{Universidade Federal do Rio de Janeiro - Campus Duque de Caxias}, \orgaddress{\city{Duque de Caxias}, \postcode{25240-005}, \state{RJ}, \country{Brazil}}}

\affil[5]{\orgdiv{Departament de F\'isica}, \orgname{Universitat Aut\`onoma de Barcelona}, \postcode{08193}, \orgaddress{\city{Bellaterra}, \country{Spain}}}

\affil[6]{\orgdiv{Independent Researcher}, \orgaddress{\city{Barcelona}, \postcode{08193}, \country{Spain}}}

\affil[7]{\orgname{International School for Advanced Studies (SISSA)}, \orgaddress{\city{Trieste}, \postcode{I-34136}, \country{Italy}}}

\singlespacing

\maketitle

\section{Geometric Diffusion}\label{geodiff}

In semiclassical transport, the diffusion tensor $D_{ij}$ is related to the integrated auto-velocity correlation 
\begin{equation}
    D_{ij} = \int_0^\infty \langle v_i(t) v_j(0) \rangle \, dt,
\end{equation}
where $v_i = \partial_{k_i} \epsilon(\mathbf{k}) / \hbar$ is the band velocity for simple bands, but can be generalized to include geometric effects. The quantum metric enhances velocity fluctuations $\delta v$ because it encodes the spread of the wavefunction in momentum space, and will  contribute an additional variance
  \begin{equation}
  \langle \delta v_i \delta v_j \rangle \sim g_{ij}(\mathbf{k}).
  \end{equation}
This arises because the quantum metric governs the overlap fluctuations between states at nearby $\mathbf{k}$-points. In systems with strong geometric effects (e.g., twisted bilayer graphene, topological insulators, or Dirac/Weyl materials), this leads to diffusion beyond the Drude model.

\subsection{Velocity auto-correlation and quantum metric}

To derive the velocity variance $\langle \delta v_i \delta v_j \rangle$ from the quantum metric $g_{ij}(\mathbf{k})$, we start from the geometric structure of Bloch states in momentum space and connect it to velocity fluctuations. The velocity operator for a Bloch electron is given by the derivative of $\mathcal{H}(\mathbf{k})$
\begin{equation}
    v_i = \frac{1}{\hbar} \partial_{k_i} \mathcal{H}(\mathbf{k}).
\end{equation}
Its matrix elements between $|u_n(\mathbf{k})\rangle$ and $|u_m(\mathbf{k})\rangle$ are
\begin{equation}
    v_i^{nm}(\mathbf{k}) = \frac{1}{\hbar} \langle u_n(\mathbf{k}) | \partial_{k_i} \mathcal{H}(\mathbf{k}) | u_m(\mathbf{k}) \rangle.
\end{equation}
For an eigenstate $|u_n(\mathbf{k})\rangle$ of $\mathcal{H}(\mathbf{k})$ with eigenvalue $\epsilon_n(\mathbf{k})$
\begin{equation}
    \partial_{k_i} \epsilon_n(\mathbf{k}) = \langle u_n(\mathbf{k}) | \partial_{k_i} \mathcal{H}(\mathbf{k}) | u_n(\mathbf{k}) \rangle.
\end{equation}
These correspond to the intraband matrix elements.

However, for interband elements ($n \neq m$), we need to extend our calculation. We start with the equation
\begin{equation}
    \mathcal{H}(\mathbf{k}) |u_m(\mathbf{k})\rangle = \epsilon_m(\mathbf{k}) |u_m(\mathbf{k})\rangle.
\end{equation}
Then we take the derivative with respect to $k_i$
\begin{equation}
    (\partial_{k_i} \mathcal{H}) |u_m\rangle + \mathcal{H} |\partial_{k_i} u_m\rangle = (\partial_{k_i} \epsilon_m) |u_m\rangle + \epsilon_m |\partial_{k_i} u_m\rangle.
\end{equation}
Next, we multiply by $\langle u_n|$ (for $n \neq m$)
\begin{equation}
    \langle u_n | \partial_{k_i} \mathcal{H} | u_m \rangle + \langle u_n | \mathcal{H} | \partial_{k_i} u_m \rangle = \epsilon_m \langle u_n | \partial_{k_i} u_m \rangle.
\end{equation}
Since $\mathcal{H} \langle u_n| = \epsilon_n \langle u_n|$, the second term becomes
\begin{equation}
    \langle u_n | \mathcal{H} | \partial_{k_i} u_m \rangle = \epsilon_n \langle u_n | \partial_{k_i} u_m \rangle.
\end{equation}
We can now substitute back to arrive at
\begin{equation}
    \langle u_n | \partial_{k_i} \mathcal{H} | u_m \rangle = (\epsilon_m - \epsilon_n) \langle u_n | \partial_{k_i} u_m \rangle.
\end{equation}
Thus, the velocity matrix element is
\begin{equation}
    v_i^{nm}(\mathbf{k}) = \frac{1}{\hbar} (\epsilon_m - \epsilon_n) \langle u_n | \partial_{k_i} u_m \rangle.
\end{equation}
The factor $(\epsilon_m - \epsilon_n)$ indicates that transitions between energetically separated bands contribute more strongly to velocity fluctuations, while the overlap $\langle u_n | \partial_{k_i} u_m \rangle$ is tied to the quantum metric $g_{ij}(\mathbf{k})$ when summed over virtual transitions.

The velocity variance $\langle \delta v_i \delta v_j \rangle$ arises from transitions to other bands and is tied to the quantum metric $g_{ij}(\mathbf{k})$. The quantum metric for a single band $n$ is defined as
\begin{equation}
    g_{ij}(\mathbf{k}) = \text{Re} \left[ \langle \partial_{k_i} u_n | \partial_{k_j} u_n \rangle - \langle \partial_{k_i} u_n | u_n \rangle \langle u_n | \partial_{k_j} u_n \rangle \right].
\end{equation}
This quantity measures the distance between nearby Bloch states in Hilbert space.
The term $ \langle \partial_{k_i} u_n | \partial_{k_j} u_n \rangle $ can be expanded using the completeness relation 
\begin{equation}
    \sum_m |u_m\rangle \langle u_m| = 1,
\end{equation}
where $ m $ runs over all bands. Now
\begin{equation}
    \langle \partial_{k_i} u_n | \partial_{k_j} u_n \rangle = \sum_m \langle \partial_{k_i} u_n | u_m \rangle \langle u_m | \partial_{k_j} u_n \rangle.
\end{equation}
This includes both the term where $ m = n $ and terms where $ m \neq n $.
The sum is now split in two parts
\begin{eqnarray}
    \langle \partial_{k_i} u_n | \partial_{k_j} u_n \rangle &=& \langle \partial_{k_i} u_n | u_n \rangle \langle u_n | \partial_{k_j} u_n \rangle \nonumber\\
    &+& \sum_{m \neq n} \langle \partial_{k_i} u_n | u_m \rangle \langle u_m | \partial_{k_j} u_n \rangle.
\end{eqnarray}
The first term $ \langle \partial_{k_i} u_n | u_n \rangle \langle u_n | \partial_{k_j} u_n \rangle $ is the product of Berry connections (phase-related terms). The second term $ \sum_{m \neq n} $ captures interband transitions. The Berry connection terms cancel out, leaving us with
\begin{equation}
    g_{ij}(\mathbf{k}) = \text{Re} \left[ \sum_{m \neq n} \langle \partial_{k_i} u_n | u_m \rangle \langle u_m | \partial_{k_j} u_n \rangle \right].
\end{equation}
This shows that the quantum metric is entirely determined by interband transitions ($ m \neq n $), reflecting the geometry of the Bloch states.

From the velocity operator expression, the off-diagonal elements ($m \neq n$) are
\begin{equation}
    v_i^{nm}(\mathbf{k}) = \frac{\epsilon_m - \epsilon_n}{\hbar} \langle u_n | \partial_{k_i} u_m \rangle.
\end{equation}
Thus, the product of velocity fluctuations is
\begin{equation}
    v_i^{nm} v_j^{mn} = \frac{\epsilon_m - \epsilon_n}{\hbar}\frac{\epsilon_n - \epsilon_m}{\hbar} \langle u_n | \partial_{k_i} u_m \rangle \langle u_m | \partial_{k_j} u_n \rangle,
\end{equation}
and $(\epsilon_m - \epsilon_n)(\epsilon_n - \epsilon_m)=-(\epsilon_m - \epsilon_n)^2$.
We can further simplify this expression by using the fact that $\langle u_n | u_m \rangle = 0$ for $n \neq m$. In this case we can write
\begin{equation}
    \partial_{k_i} \langle u_n | u_m \rangle = 0\;\;\Longrightarrow \;\; 
    \langle \partial_{k_i} u_n | u_m \rangle =-
    \langle u_n | \partial_{k_i} u_m \rangle,
\end{equation}
and the product of velocity fluctuations becomes 
\begin{equation}
    v_i^{nm} v_j^{mn} = \left( \frac{\epsilon_m - \epsilon_n}{\hbar} \right)^2 \langle \partial_{k_i} u_n | u_m \rangle \langle u_m | \partial_{k_j} u_n \rangle.
\end{equation}

\subsection{The case of a two-band Hamiltonian}

We now specialize to a general two-band Hamiltonian with eigenstates $|+\rangle$ and $|-\rangle$. The product of interband velocity matrix elements is
\begin{equation}
    v_i^{-+} v_j^{+-} = \left( \frac{\epsilon_+ - \epsilon_-}{\hbar} \right)^2 \langle \partial_{k_i} u_- | u_+ \rangle \langle u_+ | \partial_{k_j} u_- \rangle.
\end{equation}
Here $v_i^{-+}$ is the velocity matrix element from the lower band ($-$) to the upper band ($+$), $\epsilon_\pm$ are the eigenstates for the upper/lower bands, and $\langle \partial_{k_i} u_- | u_+ \rangle$ is the overlap between the $k_i$-derivative of $|u_-\rangle$ and $|u_+\rangle$.

In a two-band system, the only virtual transition contributing to the velocity variance is $(-) \leftrightarrow (+)$. Thus, the total variance is
\begin{equation}
    \langle \delta v_i \delta v_j \rangle = v_i^{-+} v_j^{+-} + v_i^{+-} v_j^{-+}.
\end{equation}
However, since $v_i^{+-} = (v_i^{-+})^*$ (Hermiticity of the velocity operator), we can write
\begin{equation}
    \langle \delta v_i \delta v_j \rangle = v_i^{-+} v_j^{+-} + \text{c.c.} = 2 \, \text{Re} \left[ v_i^{-+} v_j^{+-} \right].
\end{equation}
Substituting now
\begin{equation}
    \langle \delta v_i \delta v_j \rangle = 2 \, \text{Re} \left[ \left( \frac{\Delta \epsilon}{\hbar} \right)^2 \langle \partial_{k_i} u_- | u_+ \rangle \langle u_+ | \partial_{k_j} u_- \rangle \right],
\end{equation}
where $\Delta \epsilon = \epsilon_+ - \epsilon_-$. The quantum metric $g_{ij}(\mathbf{k})$ for the lower band ($|u_-\rangle$) is
\begin{equation}
    g_{ij}(\mathbf{k}) = \text{Re} \left[ \langle \partial_{k_i} u_- | \partial_{k_j} u_- \rangle - \langle \partial_{k_i} u_- | u_- \rangle \langle u_- | \partial_{k_j} u_- \rangle \right].
\end{equation}
For a two-band system, the completeness relation is
\begin{equation}
    |u_+\rangle \langle u_+| + |u_-\rangle \langle u_-| = 1.
\end{equation}
Thus, the first term expands as
\begin{eqnarray}
    \langle \partial_{k_i} u_- | \partial_{k_j} u_- \rangle &=& \langle \partial_{k_i} u_- | u_+ \rangle \langle u_+ | \partial_{k_j} u_- \rangle \nonumber\\
    &+& \langle \partial_{k_i} u_- | u_- \rangle \langle u_- | \partial_{k_j} u_- \rangle.
\end{eqnarray}
Substituting into $g_{ij}(\mathbf{k})$
\begin{equation}
    g_{ij}(\mathbf{k}) = \text{Re} \left[ \langle \partial_{k_i} u_- | u_+ \rangle \langle u_+ | \partial_{k_j} u_- \rangle \right].
\end{equation}
We can now combine the results
\begin{eqnarray}
    \langle \delta v_i \delta v_j \rangle &=& 2 \left( \frac{\Delta \epsilon}{\hbar} \right)^2 \text{Re} \left[ \langle \partial_{k_i} u_- | u_+ \rangle \langle u_+ | \partial_{k_j} u_- \rangle \right] \nonumber\\
    &=& 2 \left( \frac{\Delta \epsilon}{\hbar} \right)^2 g_{ij}(\mathbf{k}).
\end{eqnarray}
The total velocity variance (averaged over all virtual transitions) for a two-band model becomes simply
\begin{equation}
    \langle \delta v_i \delta v_j \rangle = 2\frac{(\Delta \epsilon)^2}{\hbar^2} g_{ij}(\mathbf{k}).
\end{equation}
The quantum metric $g_{ij}(\mathbf{k})$ thus quantifies the geometric spread of Bloch states in momentum space, leading to intrinsic velocity fluctuations that contribute to diffusion beyond the Drude picture.

\section{The Drift--Zeeman interaction}
Here we demonstrate mathematically how the Drift--Zeeman coupling arises for a Weyl particle in the presence of static electric, ${\bf E}$, and magnetic, ${\bf B}$, fields. We treat the coupling to the electromagnetic field perturbatively and obtain an effective Hamiltonian. We begin with the unperturbed Hamiltonian for a Weyl fermion 
\begin{equation}
    H_0 = \lambda{\bf k}\cdot{\boldsymbol{\sigma}},
    \label{eq:H_0}
\end{equation}
with $\lambda>0$ and eigenstates $\ket{\pm}$ satisfying $H_0\ket{\pm}=\pm\lambda k\ket{\pm}$, where $k\equiv|{\bf k}|$. We consider the coupling with the electromagnetic field and take as perturbation
\begin{equation}
    H_1=H_E+H_B\equiv-e{\bf E}\cdot{\bf r}-\frac{e\lambda}{2\hbar}({\bf B}\times{\bf r})\cdot\boldsymbol{\sigma},
    \label{eq:H_1}
\end{equation}
where we have included the electric dipole interaction and the minimal coupling to the gauge potential of a background magnetic field in the symmetric gauge, i.e., ${\bf k}\rightarrow {\bf k}-e/\hbar{\bf A}$ with ${\bf A}=({\bf B}\times{\bf r})/2$.

For our two-band model, the second-order corrections to the energy levels of $H_0$ are given by
\begin{equation}
    \Delta E_{\pm}^{(2)}=\pm\frac{|\bra{\pm} H_1\ket{\mp}|^2}{\Delta E}, \qquad \Delta E\equiv E_+-E_- = 2\lambda k.
\end{equation}
For simplicity, we focus on the correction to $E_+$, and isolate the mixed contributions arising from the interplay between $H_E$ and $H_B$, which result from the term
\begin{equation}
    \Delta E_+^{(2)}=\frac{2}{\Delta E}\text{Re}[\bra{-}H_E\ket{+}\bra{+}H_B\ket{-}],
    \label{eq:mixed_pert}
\end{equation}
where we used hermiticity to write $\bra{+}H_i\ket{-}=(\bra{-}H_i\ket{+})^*$, with $H_i=H_E,H_B$. Our final goal is to show that the mixed electric-magnetic contribution generates an effective coupling proportional to $({\bf E}\times {\bf B})\cdot \boldsymbol{\sigma}$.

\subsection{Computation of the matrix elements}
\subsubsection{Electric perturbation}
The matrix element of the electric part reads $\bra{-}H_E\ket{+}=-e E_i\bra{-}r_i\ket{+}$, where we use the Einstein convention to sum over repeated indices, unless stated otherwise.
A useful commutator to evaluate is
\begin{equation}
    [r_i,H_0]=\lambda[r_i,k_j\sigma_j]=i\lambda\sigma_i\quad\Rightarrow\quad [{\bf r}, H_0]=i\lambda\boldsymbol{\sigma},
\end{equation}
therefore
\begin{equation}
    \bra{-}[r_i,H_0]\ket{+}=\Delta E\bra{-}r_i\ket{+}=i\lambda\bra{-}\sigma_i\ket{+}\quad\Rightarrow\quad \bra{-}{\bf r}\ket{+}=\frac{i\lambda}{\Delta E}\bra{-}\boldsymbol{\sigma}\ket{+},
\end{equation}
and this leads to
\begin{equation}
    \bra{-}H_E\ket{+} = -i\frac{\lambda}{\Delta E}{\bf E}\cdot\bra{-}\boldsymbol{\sigma}\ket{+}.
    \label{eq:HE_element}
\end{equation}

\subsubsection{Magnetic perturbation}
We now evaluate the magnetic part
\begin{equation}
    \bra{+}H_B\ket{-}=-\epsilon_{ijk}B_j\frac{e\lambda}{2\hbar}\bra{+}r_k\sigma_i\ket{-}.
\end{equation}
We first evaluate the following useful commutators, which will be used below:
\begin{equation}
    [r_k,H_0\sigma_i]=i\lambda\sigma_k\sigma_i,\qquad [H_0,\sigma_i]=2i\lambda k_j\epsilon_{jil}\sigma_l,
\end{equation}
from which we can write
\begin{equation}
    \bra{+}[r_k,H_0\sigma_i]\ket{-}=2i\lambda\epsilon_{jil}\bra{+}r_k k_j \sigma_l\ket{-}-\Delta E\bra{+}\sigma_ir_k\ket{-}.
\end{equation}
Solving for $\bra{+}\sigma_ir_k\ket{-}$ we get
\begin{align}
    \nonumber
    \bra{+}\sigma_ir_k\ket{-} &= \frac{1}{\Delta E}\bigg\{2i\lambda\epsilon_{jil}\bra{+}r_k k_j \sigma_l\ket{-} - \bra{+}\underbrace{[r_k,H_0\sigma_i]}_{=i\lambda\sigma_k\sigma_i}\ket{-}\bigg\}\\
    & = \frac{\lambda}{\Delta E}\epsilon_{kip}\bra{+}\sigma_p\ket{-}+i\frac{2\lambda}{\Delta E}\epsilon_{jil}\bra{+}r_k k_j \sigma_l\ket{-}
\end{align} 
Inserting this last line in the magnetic Hamiltonian matrix element, we find
\begin{align}
    \nonumber
    \bra{+}H_B\ket{-}=&-\frac{e\lambda^2}{\hbar\Delta E}\bigg\{{\bf B}\cdot\bra{+}\boldsymbol{\sigma}\ket{-}+i\;B_j(\delta_{jl}\delta_{mk}-\delta_{jm}\delta_{kl})\bra{+}r_k k_m \sigma_l\ket{-}\bigg\},
\end{align}
where we used the identities $\epsilon_{ijk}\epsilon_{kip}=2\delta_{jp}$, $\epsilon_{ijk}\epsilon_{mil}=-\epsilon_{jki}\epsilon_{mli}=\delta_{jl}\delta_{mk}-\delta_{jm}\delta_{kl}$. To express the matrix element more compactly, we rewrite the second term using fundamental tensor identities. In the right-hand side we can identify the terms
\begin{equation}
    B_j\delta_{jl}\delta_{mk}\bra{+}r_k k_m \sigma_l\ket{-}=B_j\bra{+}r_m k_m \sigma_j\ket{-}=\bra{+}({\bf r}\cdot{\bf k})({\bf B}\cdot\boldsymbol{\sigma})\ket{-},
\end{equation}
\begin{equation}
    B_j\delta_{jm}\delta_{kl}\bra{+}r_k k_m \sigma_l\ket{-} = B_j \bra{+}r_l k_j \sigma_l\ket{-} = \bra{+}({\bf r}\cdot\boldsymbol{\sigma})({\bf B}\cdot{\bf k})\ket{-},
\end{equation}
which can be also reorganized as follows. First notice that
\begin{equation}
    B_i\sigma_ir_jk_j - r_k\sigma_k k_l B_l = B_i\sigma_j (\delta_{ij}r_l k_l - r_j k_i),
    \label{eq:extra_term_manipulation}
\end{equation}
then rewrite
\begin{equation}
    B_i(\delta_{ij}r_l k_l - r_j k_i) = (B_jr_l - B_l r_j)k_l = \epsilon_{jlm} ({\bf B}\times {\bf r})_m k_l ,
\end{equation}
and replace in Eq. \eqref{eq:extra_term_manipulation} to get
\begin{equation}
    B_i\sigma_ir_jk_j - r_k\sigma_k k_l B_l = \sigma_j \epsilon_{jlm} ({\bf B}\times {\bf r})_m k_l = ({\bf B}\times {\bf r})_m ({\bf k}\times \boldsymbol{\sigma})_m = ({\bf B}\times {\bf r})\cdot ({\bf k}\times \boldsymbol{\sigma}).
\end{equation}

Therefore, the matrix element of the magnetic Hamiltonian is  
\begin{equation}
    \bra{+}H_B\ket{-} = -\frac{e\lambda^2}{\hbar\Delta E}\bigg\{{\bf B}\cdot\bra{+}{\boldsymbol{\sigma} }\ket{-}+i\;\bra{+}({\bf B}\times {\bf r})\cdot ({\bf k}\times {\boldsymbol{\sigma}})\ket{-}\bigg\}
    \label{eq:HB_element}.
\end{equation}

\subsection{Mixed term evaluation}
We can now evaluate Eq. \eqref{eq:mixed_pert} using \eqref{eq:HE_element} and \eqref{eq:HB_element}:
\begin{align}
    \nonumber
    \Delta E_+^{(2)} = \frac{2\lambda^3 e}{\hbar (\Delta E)^3}\text{Re}\bigg\{&i {\bf E}\cdot\bra{-}\boldsymbol{\sigma}\ket{+} \bigg[{\bf B}\cdot\bra{+}\boldsymbol{\sigma}\ket{-}+i\;\bra{+}({\bf B}\times {\bf r})\cdot ({\bf k}\times \boldsymbol{\sigma})\ket{-}\bigg]\bigg\}.
\end{align}

We use the completeness relation to write $\ket{+}\bra{+}=\mathbb{I}_2-\ket{-}\bra{-}$. This step is complemented by the observation that $\bra{-}\boldsymbol{\sigma}\ket{-}\in\mathbb{R}$, which allows us to simplify the evaluation of the corresponding terms\footnote{This can be seen by parametrizing 
    \begin{equation}
        \ket{+}=
        \begin{pmatrix}
            \cos\frac{\theta}{2}\\
            e^{i\phi}\sin\frac{\theta}{2}
        \end{pmatrix},\qquad
        \ket{-}=
        \begin{pmatrix}
            -\sin\frac{\theta}{2}\\
            e^{i\phi}\cos\frac{\theta}{2}
        \end{pmatrix}
    \end{equation}
    and computing explicitly the matrix element, ending up in
    \begin{equation}
        \bra{-}\boldsymbol{\sigma}\ket{-} =-
        \begin{pmatrix}
            \cos\phi\sin\theta\\
            \sin\theta\sin\phi\\
            \cos\theta
        \end{pmatrix}=-\bra{+}\boldsymbol{\sigma}\ket{+}.
\end{equation}}. In fact, we can directly replace $\ket{+}\bra{+}$ with the identity in the first term, because the term we add with this replacement involves $i\bra{-}\sigma_i\ket{-}\bra{-}\sigma_j\ket{-}$, which is purely imaginary, and as we take the real part its actual contribution is zero. Then, we rewrite the energy correction as:
\begin{align}
    \Delta E_+^{(2)}=\frac{2e\lambda^3}{\hbar(\Delta E)^3}\text{Re}\bigg\{&i\bra{-}({\bf E}\cdot\boldsymbol{\sigma})({\bf B}\cdot\boldsymbol{\sigma})\ket{-}+\bra{-}{\bf E}\cdot\boldsymbol{\sigma}\ket{+}\bra{+}({\bf B}\times {\bf r})\cdot ({\bf k}\times \boldsymbol{\sigma})\ket{-}\bigg\}.
\end{align}
By manipulating the product in the first addend, we can write it as
\begin{equation}
    ({\bf E}\cdot\boldsymbol{\sigma})({\bf B}\cdot\boldsymbol{\sigma}) = E_i B_j \sigma_i\sigma_j = E_iB_j (\delta_{ij}+i\epsilon_{ijk}\sigma_k) = {\bf E}\cdot{\bf B}+i({\bf E}\times {\bf B})\cdot\boldsymbol{\sigma}
\end{equation}
Taking the real part we eliminate the first term, $\text{Re}[i({\bf E}\cdot{\bf B}+i({\bf E}\times {\bf B})\cdot\boldsymbol{\sigma})]=-({\bf E}\times {\bf B})\cdot\boldsymbol{\sigma}$. Therefore
\begin{align}
    \Delta E_+^{(2)}=&-\frac{2e\lambda^3}{\hbar\Delta E^3}\bra{-}({\bf E}\times {\bf B})\cdot\boldsymbol{\sigma}\ket{-}+\nonumber\\&-\frac{2e\lambda^3}{\hbar\Delta E^3}\text{Re}\bigg\{\bra{-}{\bf E}\cdot\boldsymbol{\sigma}\ket{+}\bra{+}({\bf B}\times {\bf r})\cdot ({\bf k}\times \boldsymbol{\sigma})\ket{-}\bigg\}.\label{eq:sec_term_to_neglect}
\end{align}
Now, the second term in Eq. \eqref{eq:sec_term_to_neglect} should be neglected because it is not translationally invariant nor gauge invariant, since it is explicitly position dependent as well as written in terms of the gauge potential ${\bf A}\propto({\bf B}\times {\bf r})$, therefore it changes under a gauge transformation. Because both translational invariance and gauge invariance are symmetries of the original Hamiltonian, the term in \eqref{eq:sec_term_to_neglect} merely reflects the truncation inherent in the perturbative expansion, while the full perturbative series is necessarily gauge invariant.

We are left with the first term of Eq. \eqref{eq:sec_term_to_neglect}, which includes a term exactly of the form $({\bf E}\times {\bf B})\cdot\boldsymbol{\sigma}$. If the chemical potential is not too close to the Weyl node, the denominator $\Delta E^3$ can be replaced by an average energy $\Delta^3$ given, for example, by the actual bandwidth of Te. In this case, and given that the calculations to correct $E_-$ yield $\Delta E_-^{(2)}=\Delta E_+^{(2)}$, we conclude that the perturbative treatment of the electromagnetic Hamiltonian leads to
\begin{equation}
    \Delta H_{\text{eff}} = -\gamma \, (\mathbf{E} \times \mathbf{B}) \cdot \boldsymbol{\sigma},
\end{equation}
with $\gamma = \frac{2e^2 \lambda^3}{\hbar \Delta^3}$. This term has the form of an effective Zeeman interaction generated by the drift velocity, hence the name "Drift--Zeeman coupling".

Before proceeding, we should remark at this point that if we include in the full Hamiltonian, $H_0$, also the kinetic term $\propto \alpha k^2 \mathbb{I}_2$, and if we repeat the same perturbative calculation to account for the minimal coupling, $k^2\to \vert\mathbf{k}-\mathbf{A}\vert^2$, we obtain an extra contribution to $\Delta H_{\mathrm{eff}}$ of the type $\kappa\; \mathbf{w}(\mathbf{E},\mathbf{B},\mathbf{k})\cdot\mathbf{k}\;\mathbb{I}_2$, where $\kappa= 2e\lambda^2 \alpha/2\hbar\Delta^3$ and $\mathbf{w}$ is a rather involved expression of the electromagnetic field and the momentum. Since this term simply leads to a shift of the Bloch velocity, and does not contribute to the spin splitting, it will therefore be neglected in the following calculations.

\subsection{Band splitting induced by magnetoelectric spin coupling}

We now introduce a spin-orbit interaction induced by the combined effect of an external magnetic field and the internal electric polarization. The magnetic field acts on the orbital motion of the electrons, and this orbital motion is coupled to the spin through the spin-orbit interaction, which is in turn modified by the presence of an internal electric polarization. 

The anisotropic Hamiltonian is
\begin{equation}
H = \frac{\hbar^2 k_\perp^2}{2m_\perp^*} + \frac{\hbar^2 k_z^2}{2m_z^*} + \lambda_\perp (k_x \sigma_x + k_y \sigma_y) + \lambda_z k_z \sigma_z - \gamma\cdot (\boldsymbol{\cal E}\times{\bf B}) \cdot \boldsymbol{\sigma},
\end{equation}
where $k_\perp^2 = k_x^2 + k_y^2$, $\boldsymbol{\sigma} = (\sigma_x, \sigma_y, \sigma_z)$ are the Pauli matrices, $\boldsymbol{\cal E} = ({\cal E}_x, {\cal E}_y, {\cal E}_z)$ is the electric field associated to the polarization due to lone pairs, ${\bf B} = (B_x, B_y, B_z)$ is the applied magnetic field. The parameters $\gamma_z=2e^2\lambda_z^3/\hbar\Delta^3$ and $\gamma_\perp=2e^2\lambda_\perp^3/\hbar\Delta^3$ are the anisotropic Drift--Zeeman coupling constants, such that $\gamma_z>\gamma_\perp$ in tellurene. 
The Hamiltonian can be rewritten in a compact form
\begin{equation}
H = H_0 \mathbb{I}_2 + \mathbf{d} \cdot \vec{\sigma},
\end{equation}
where $H_0 = \frac{\hbar^2 k_\perp^2}{2m_\perp^*} + \frac{\hbar^2 k_z^2}{2m_z^*}$ is the scalar part, $\mathbf{d} = (d_x, d_y, d_z)$ is the effective magnetic field due to spin-orbit coupling and the $\boldsymbol{\cal E}\times{\bf B}$ term. The $ \mathbf{d} $ vector has two contributions
\begin{equation}
   \mathbf{d}_{\text{SO}} = (\lambda_\perp k_x, \lambda_\perp k_y, \lambda_z k_z),
\end{equation}
and a magnetoelectric contribution, such that
\begin{equation}
    \mathbf{d} = \mathbf{d}_{\text{SO}} - \gamma\cdot (\boldsymbol{\cal E}\times{\bf B}).
\end{equation}
Explicitly
\begin{eqnarray}
d_x &=& \lambda_\perp k_x - \gamma_\perp ({\cal E}_y B_z-{\cal E}_z B_y), \nonumber \\
d_y &=& \lambda_\perp k_y - \gamma_\perp ({\cal E}_z B_x-{\cal E}_x B_z), \nonumber \\
d_z &=& \lambda_z k_z - \gamma_z ({\cal E}_x B_y-{\cal E}_y B_x).\nonumber
\end{eqnarray}
The eigenvalues of $ H $ are given by $E_\pm = H_0 \pm |\mathbf{d}|$, where $|\mathbf{d}| = \sqrt{d_x^2 + d_y^2 + d_z^2}$.
Thus
\begin{equation}
    E_\pm = \frac{\hbar^2 k_\perp^2}{2m_\perp^*} + \frac{\hbar^2 k_z^2}{2m_z^*} \pm \sqrt{d_x^2 + d_y^2 + d_z^2}.
\end{equation}
The band splitting is given by the difference between the two eigenvalues,
\begin{equation}
    \Delta E = E_+ - E_- = 2 |\mathbf{d}| = 2 \sqrt{d_x^2 + d_y^2 + d_z^2}.
\end{equation}
The square of the splitting is
\begin{equation}
    (\Delta E)^2 = 4 (d_x^2 + d_y^2 + d_z^2).
\end{equation}
Substituting $ d_x, d_y, d_z $
\begin{eqnarray}
    (\Delta E)^2 &=& 4 \left[ (\lambda_\perp k_x - \gamma_\perp ({\cal E}_y B_z-{\cal E}_z B_y))^2 + (\lambda_\perp k_y - \gamma_\perp ({\cal E}_z B_x-{\cal E}_x B_z))^2
    \right.\nonumber\\
    &+& \left. (\lambda_z k_z - \gamma_z ({\cal E}_x B_y-{\cal E}_y B_x))^2 \right].
\end{eqnarray}
If we consider a two-dimensional (2D) system confined to the $x-z $ plane (i.e., $k_y = 0$), and for an electric field along the $x-$direction, $\boldsymbol{\cal E}\approx({\cal E}_x,0,0)$, the expression for $(\Delta E)^2$ simplifies significantly
\begin{equation}
(\Delta E)^2 = 4 \left[
\lambda_\perp^2 k_x^2 + \lambda_z^2 k_z^2 
- 2 \gamma_z {\cal E}_x \lambda_z k_z B_y 
+ {\cal E}_x^2 (\gamma_z^2 B_y^2 + \gamma_\perp^2 B_z^2)
\right].
\end{equation}
For later reference we also report the analytic expressions for the zero-field components of the quantum metric tensor:
\begin{eqnarray}
g_{zz}({\bf k})&=&\frac{\lambda_z^2 \lambda_\perp^2 k_x^2}{4 \left[k_x^2\lambda_\perp^2+k_z^2\lambda_z^2\right]^2},\nonumber\\
g_{zx}({\bf k})&=&-\frac{\lambda_z^2 \lambda_\perp^2 k_x k_z}{4 \left[k_x^2\lambda_\perp^2+k_z^2\lambda_z^2\right]^2}.
\end{eqnarray}

\section{Fitting Formulae}

\subsection{Angular dependence}

We begin from the Kubo-Greenwood formula for the conductivity tensor in an $n$-band system
\begin{equation}
\sigma_{ij} = e^{2} \sum_{n} \int \frac{d^{d}\mathbf{k}}{(2\pi)^{d}} D^{(n)}_{ij}(\mathbf{k}) \left(-\frac{\partial f(\epsilon_{n}(\mathbf{k}))}{\partial \epsilon_{n}(\mathbf{k})} \right),
\label{eq:kubo}
\end{equation}
where $D^{(n)}_{ij}(\mathbf{k})$ is the diffusion tensor. At low temperatures, $-\partial f/\partial \epsilon \approx \delta(\epsilon - \mu)$, which simplifies Eq. \eqref{eq:kubo} to
\begin{equation}
\sigma_{ij} \approx e^{2} N(E_F) D_{ij}(\mu),
\label{eq:sigma_simple}
\end{equation}
where $N(E_F)$ is the density of states at the Fermi level.
The total diffusion tensor $D_{ij}(\mu)$ consists of a conventional Drude contribution and a quantum geometric contribution,
\begin{equation}
D_{ij}(\mu) = \underbrace{D^{D}_{ij}(\mu)}_{\text{intraband}} + \underbrace{\frac{2\tau_g}{\hbar^{2}} \langle [\Delta\epsilon(\mathbf{k})]^{2} g_{ij}(\mathbf{k}) \rangle_{FS}}_{\text{interband (geometric)}},
\label{eq:diffusion_total}
\end{equation}
where $\tau_g$ is the quantum geometric relaxation time and $g_{ij}(\mathbf{k})$ is the quantum metric.
For the $zz$-component relevant to our longitudinal resistance measurements
\begin{equation}
D_{zz}(\mu) = D^{D}_{zz}(\mu) + \frac{2\tau_g}{\hbar^{2}} \langle [\Delta\epsilon(\mathbf{k})]^{2} g_{zz}(\mathbf{k}) \rangle_{FS}.
\label{eq:D_zz}
\end{equation}
The Fermi surface average expands as
\begin{align}
\langle [\Delta \epsilon(\mathbf{k})]^{2} g_{zz}(\mathbf{k})\rangle_{FS} &= -8\underbrace{\gamma_z \langle k_z g_{zz}(\mathbf{k})\rangle_{FS} \, \hat{z} \cdot (\boldsymbol{\mathcal{E}} \times \mathbf{B})}_{\textrm{antisymmetric}} \nonumber \\
&+ \underbrace{4 \langle (\lambda_\perp^{2} k_x^{2} + \lambda_z^{2} k_z^{2}) g_{zz}(\mathbf{k})\rangle_{FS}}_{ \textrm{symmetric, $B$-independent}} \nonumber \\
&+ \underbrace{4\langle g_{zz}(\mathbf{k})\rangle_{FS} \, [\gamma \cdot (\boldsymbol{\mathcal{E}} \times \mathbf{B})]^{2}}_{\textrm{symmetric, $B^{2}$-dependent}},
\label{eq:FS_expansion}
\end{align}
given our experimental geometry ($\boldsymbol{\mathcal{E}} \parallel \hat{x}$, $\mathbf{B} \parallel \hat{y}$, transport measurements along $\hat{z}$)
\begin{equation}
\hat{z} \cdot (\boldsymbol{\mathcal{E}} \times \mathbf{B}) = \mathcal{E}_x B_y, \quad \gamma \cdot (\boldsymbol{\mathcal{E}} \times \mathbf{B}) = \gamma \mathcal{E}_x B_y.
\end{equation}
The symmetric-in-$B$ part of $D_{zz}$ becomes
\begin{equation}
D_{zz}^{\text{sym}}(B) = D^{D}_{zz} + \frac{2\tau_g}{\hbar^{2}} S_0 + \frac{2\tau_g}{\hbar^{2}} 4\langle g_{zz}(\mathbf{k})\rangle_{FS} \gamma^{2} \mathcal{E}_x^{2} B_y^{2},
\label{eq:Dzz_sym}
\end{equation}
where $S_0 = 4\langle (\lambda_\perp^{2} k_x^{2} + \lambda_z^{2} k_z^{2}) g_{zz}(\mathbf{k})\rangle_{FS}$.
The conductivity is then
\begin{equation}
\sigma_{zz}(B) = e^{2} N(E_F) D_{zz}^{\text{sym}}(B) = \sigma_0 + \alpha B_y^{2},
\label{eq:sigma_B}
\end{equation}
with
\begin{align}
\sigma_0 &= e^{2} N(E_F) \left( D^{D}_{zz} + \frac{2\tau_g}{\hbar^{2}} S_0 \right), \\
\alpha &= e^{2} N(E_F) \frac{8\tau_g}{\hbar^{2}} \langle g_{zz}(\mathbf{k})\rangle_{FS} \gamma^{2} \mathcal{E}_x^{2}.
\end{align}
The resistance $R_{zz} = \frac{1}{\sigma_{zz}} \frac{L}{A}$ expands for weak magnetic fields as
\begin{equation}
R(B) \approx \frac{L}{A} \left[ \frac{1}{\sigma_0} - \frac{\alpha}{\sigma_0^{2}} B_y^{2} + O(B^{4}) \right],
\end{equation}
yielding the fitting form
\begin{equation}
R(B) = R_0 - C B_y^{2},
\end{equation}
where
\begin{equation}
C = \frac{L}{A} \frac{\alpha}{\sigma_0^{2}}.
\label{eq:C0_general}
\end{equation}
Substituting for $\alpha$ and $\sigma_0$
\begin{equation}
C = \frac{L}{A} \frac{ e^{2} N(E_F) \frac{8\tau_g}{\hbar^{2}} \langle g_{zz} \rangle_{FS} \gamma^{2} \mathcal{E}_x^{2} }{ \left[ e^{2} N(E_F) \left( D^D_{zz} + \frac{2\tau_g}{\hbar^{2}} S_0 \right) \right]^{2} }.
\end{equation}
Using the relationship between diffusion and conductivity, $\sigma=e^2N(E_F)D$, we find
\begin{equation}
C = \frac{(L/A)e^{2} N(E_F)(8\tau_g/\hbar^{2}) \langle g_{zz} \rangle_{FS} \gamma^{2} \mathcal{E}_x^{2} }{ \left( \sigma^D_{zz} + \sigma^g_{zz} \right)^{2} },
\label{eq:C0_final_general}
\end{equation}
where the ${\bf B}-$independent, quantum geometric conductivity $\sigma^g_{zz}=N(E_F)(2\tau_g/\hbar^{2})S_0$. For the general case in which the magnetic field is rotated away from the $y-$axis,
\begin{equation}
C(\theta)=\frac{(L/A)\; e^2N(E_F) \; (8\tau_g/\hbar^{2}) \; \langle g_{zz}({\bf k}) \rangle_{FS}\;\mathcal{E}_x^{2}}{ \left( \sigma_{zz}^D + \sigma_{zz}^{g}\right)^{2} }(\gamma_z^{2} \cos^2\theta_{(\mathbf{B},\hat{y})}+\gamma_\perp^2\cos^2\theta_{(\mathbf{B},\hat{z})} ),\nonumber
\end{equation}
where $\theta_{(\mathbf{B},\hat{y})}$ and $\theta_{(\mathbf{B},\hat{z})}$ are, respectively, the angles between the magnetic field ${\bf B}$ and the $y-$ and $z-$axis.
Finally, assuming $\sigma^{D}_{zz} \gg \sigma^{g}_{zz}$
\begin{equation}
C(\theta) \approx \frac{(L/A)\; e^2N(E_F) \; (8\tau_g/\hbar^{2}) \; \langle g_{zz} \rangle_{FS} \; \mathcal{E}_x^{2}}{ (\sigma^{D}_{zz})^{2} }(\gamma_z^{2} \cos^2\theta_{(\mathbf{B},\hat{y})}+\gamma_\perp^2\cos^2\theta_{(\mathbf{B},\hat{z})} ).
\label{eq:C0_approx}
\end{equation}

\subsection{Gate voltage dependence}

For a 2D parabolic band, the relevant physical quantities scale with the carrier density \(n\) as follows: the Fermi wavevector scales as \(k_F = (2\pi n)^{1/2}\), the density of states at the Fermi level is constant, \(N(E_F) = m^*/(\pi\hbar^2)\), the Fermi-surface averaged quantum metric scales as \(\langle g_{zz} \rangle_{FS} \sim 1/k_F \sim 1/\sqrt{n}\) after integration over the 2D Fermi surface, and the Drude diffusion coefficient scales linearly with density, \(D_{zz}^D \propto n\). Substituting these scaling relations into Eq. (89) yields the gate‑dependence of the curvature parameter
\begin{equation}
C(n) \propto \frac{1}{N(E_F)} \cdot \frac{1/\sqrt{n}}{n^2} \propto n^{-5/2}.
\end{equation}
This result captures the strong decay of the negative magnetoresistance with increasing carrier density, in agreement with experimental observations.

\section{Fitting procedure}

The magnetoresistance data for different magnetic field orientations were fitted using a modified least-squares procedure implemented in Mathematica. For each angle $\theta$, the corresponding dataset was extracted from the experimental measurements and downsampled to approximately 50 points to improve numerical stability. A custom magnetic field range of $B = 1$--$8$ T was applied to all datasets to ensure consistent comparison. 

The fitting model employed was $R(B) = A_0 + F_0|B| - C_0 B^2$, where the absolute value term $F_0|B|$ accounts for possible antisymmetric contributions to the magnetoresistance. The parameters $A_0$, $C_0$, and $F_0$ were obtained via pseudoinverse linear regression applied to the design matrix $X = [1, |B|, -B^2]$. 

To ensure the robustness of the fit, quality filters were applied, excluding datasets for which $A_0 \notin [1500, 1850]~\Omega$, $C_0 \notin (-1, 5)~\Omega$/T$^2$, or $|F_0| > 20~\Omega$/T. The curvature parameter $C_0(\theta)$ was then analyzed as a function of the magnetic field orientation. Angular averages were computed for parallel ($\theta \leq 30^\circ$ or $\geq 150^\circ$) and perpendicular ($60^\circ \leq \theta \leq 120^\circ$) orientations to extract $C_z$ and $C_\perp$, respectively. Finally, these parameters were fitted to a $\cos(2\theta)$ dependence, $C(\theta) = C_{A0} + C_{B0}\cos(2\theta)$, providing a phenomenological description of the angular dependence of the magnetoresistance curvature.

\begin{figure}[htbp]
    \centering
    \begin{subfigure}[b]{0.24\textwidth}
        \centering
        \includegraphics[width=\linewidth]{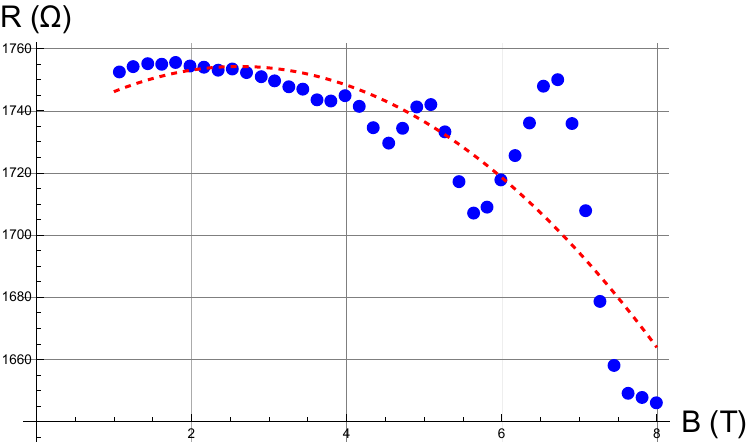}
        \caption{$\theta = 173.9^\circ$}
        \label{fig:angle173.9}
    \end{subfigure}
    \hfill
    \begin{subfigure}[b]{0.24\textwidth}
        \centering
        \includegraphics[width=\linewidth]{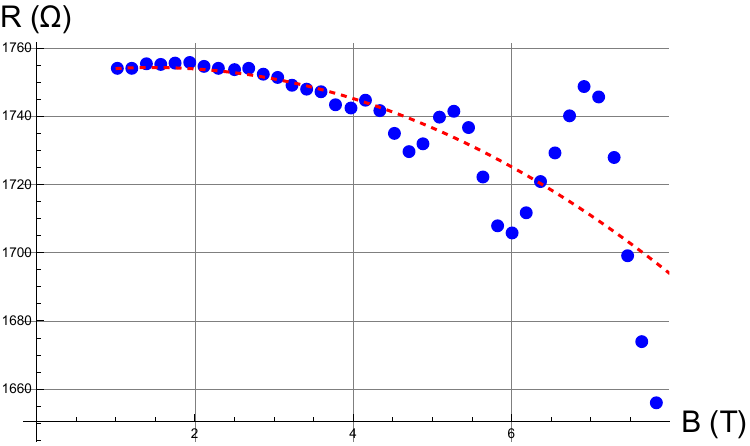}
        \caption{$\theta = 160.6^\circ$}
        \label{fig:angle160.6}
    \end{subfigure}
    \hfill
    \begin{subfigure}[b]{0.24\textwidth}
        \centering
        \includegraphics[width=\linewidth]{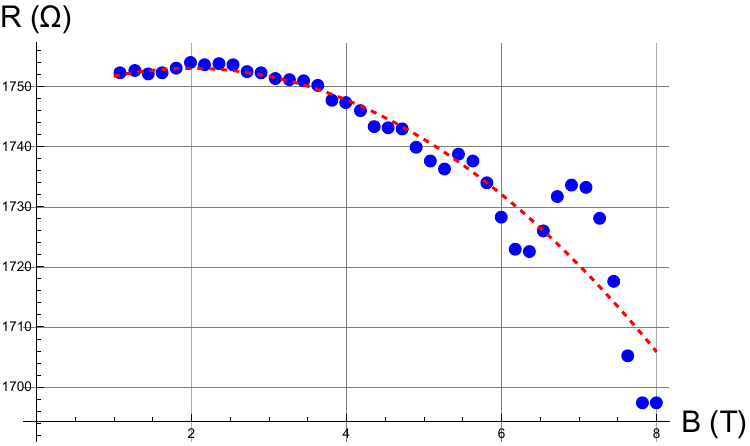}
        \caption{$\theta = 134.1^\circ$}
        \label{fig:angle134.1}
    \end{subfigure}
    \hfill
    \begin{subfigure}[b]{0.24\textwidth}
        \centering
        \includegraphics[width=\linewidth]{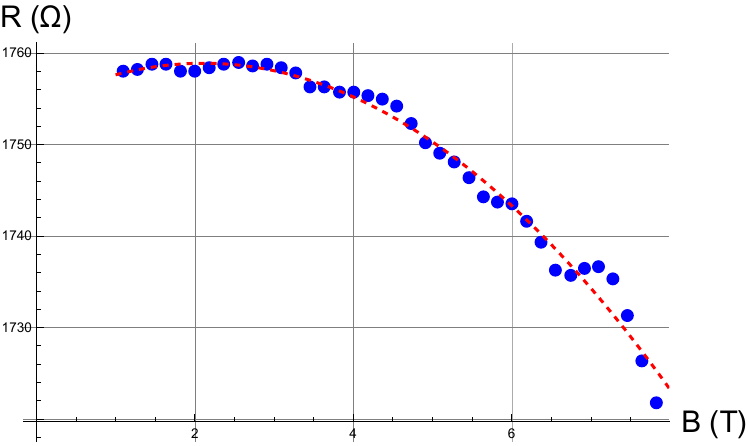}
        \caption{$\theta = 122.3^\circ$}
        \label{fig:angle122.3}
    \end{subfigure}
    
    \vspace{0.5cm}
    \begin{subfigure}[b]{0.24\textwidth}
        \centering
        \includegraphics[width=\linewidth]{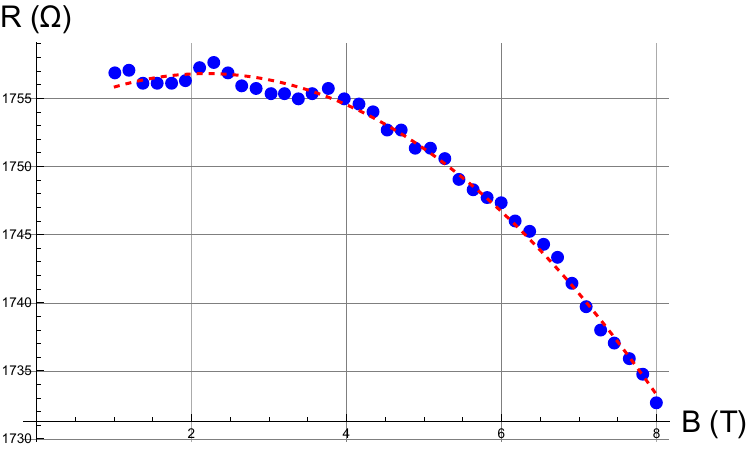}
        \caption{$\theta = 110.5^\circ$}
        \label{fig:angle110.5}
    \end{subfigure}
    \hfill
    \begin{subfigure}[b]{0.24\textwidth}
        \centering
        \includegraphics[width=\linewidth]{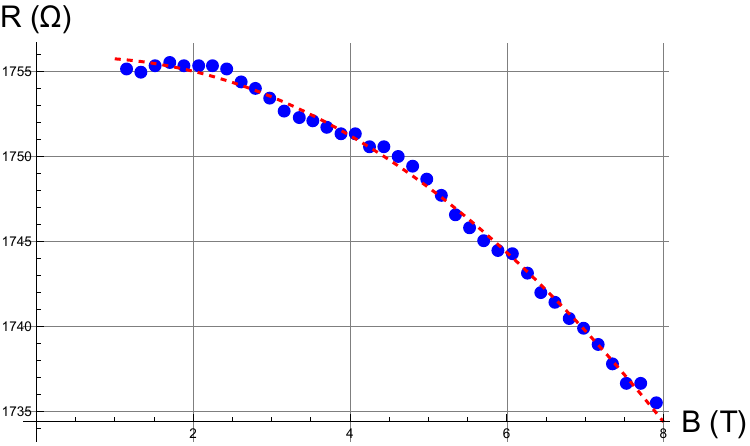}
        \caption{$\theta = 98.7^\circ$}
        \label{fig:angle98.7}
    \end{subfigure}
    \hfill
    \begin{subfigure}[b]{0.24\textwidth}
        \centering
        \includegraphics[width=\linewidth]{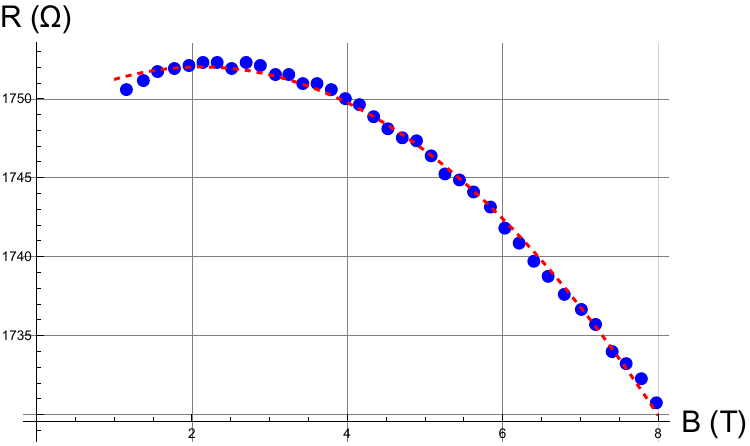}
        \caption{$\theta = 88.4^\circ$}
        \label{fig:angle88.4}
    \end{subfigure}
    \hfill
    \begin{subfigure}[b]{0.24\textwidth}
        \centering
        \includegraphics[width=\linewidth]{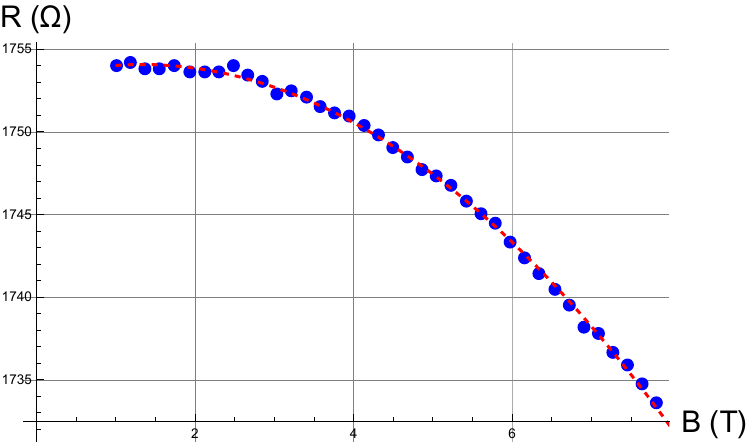}
        \caption{$\theta = 82.5^\circ$}
        \label{fig:angle82.5}
    \end{subfigure}
    
    \vspace{0.5cm}
    \begin{subfigure}[b]{0.24\textwidth}
        \centering
        \includegraphics[width=\linewidth]{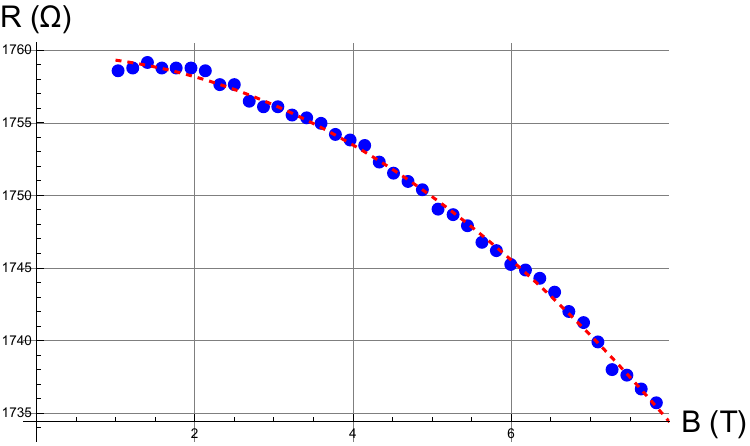}
        \caption{$\theta = 75.2^\circ$}
        \label{fig:angle75.2}
    \end{subfigure}
    \hfill
    \begin{subfigure}[b]{0.24\textwidth}
        \centering
        \includegraphics[width=\linewidth]{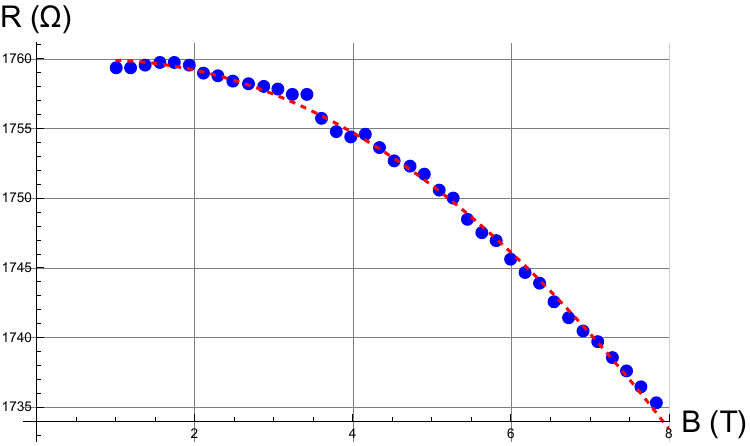}
        \caption{$\theta = 67.8^\circ$}
        \label{fig:angle67.8}
    \end{subfigure}
    \hfill
    \begin{subfigure}[b]{0.24\textwidth}
        \centering
        \includegraphics[width=\linewidth]{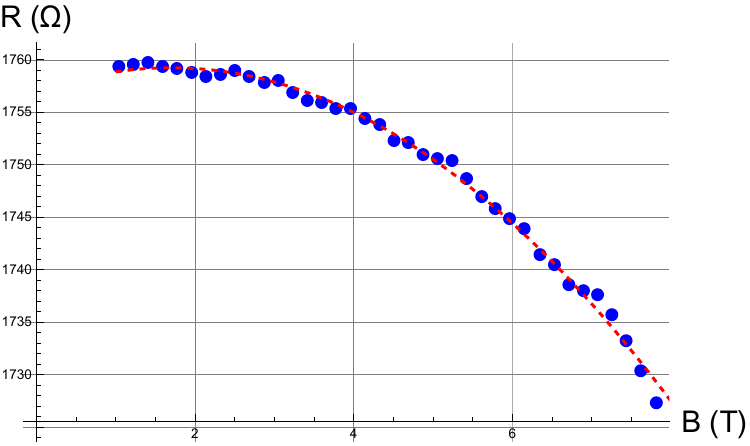}
        \caption{$\theta = 60.4^\circ$}
        \label{fig:angle60.4}
    \end{subfigure}
    \hfill
    \begin{subfigure}[b]{0.24\textwidth}
        \centering
        \includegraphics[width=\linewidth]{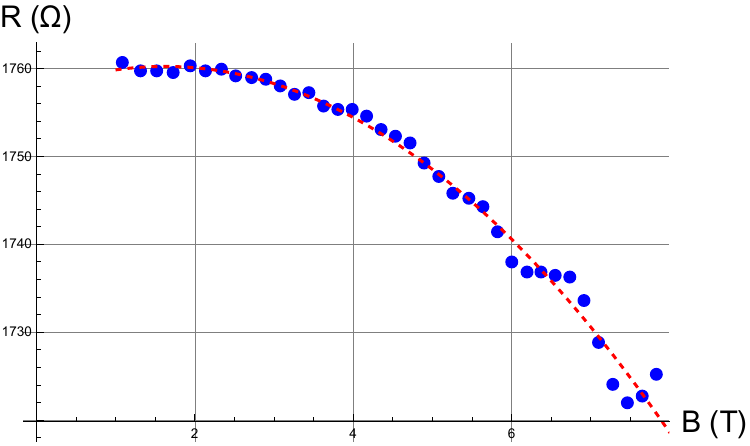}
        \caption{$\theta = 51.6^\circ$}
        \label{fig:angle51.6}
    \end{subfigure}
    
    \vspace{0.5cm}
    \begin{subfigure}[b]{0.24\textwidth}
        \centering
        \includegraphics[width=\linewidth]{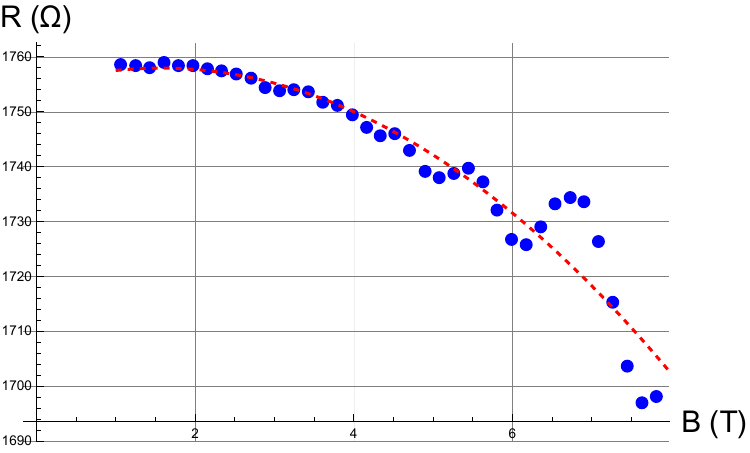}
        \caption{$\theta = 41.3^\circ$}
        \label{fig:angle41.3}
    \end{subfigure}
    \hfill
    \begin{subfigure}[b]{0.24\textwidth}
        \centering
        \includegraphics[width=\linewidth]{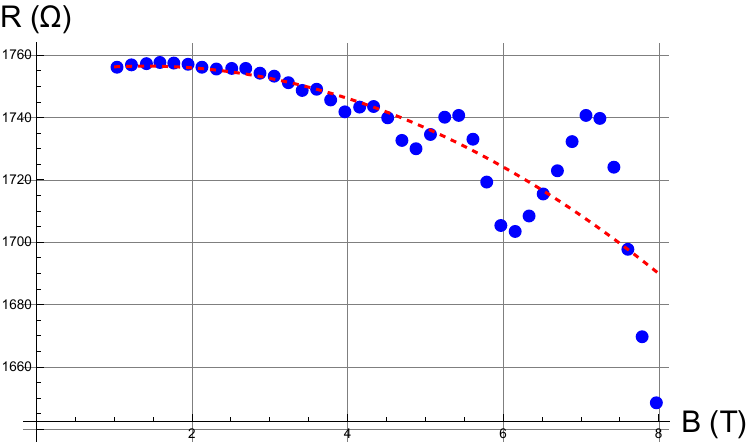}
        \caption{$\theta = 20.6^\circ$}
        \label{fig:angle20.6}
    \end{subfigure}
    \hfill
    \begin{subfigure}[b]{0.24\textwidth}
        \centering
        \includegraphics[width=\linewidth]{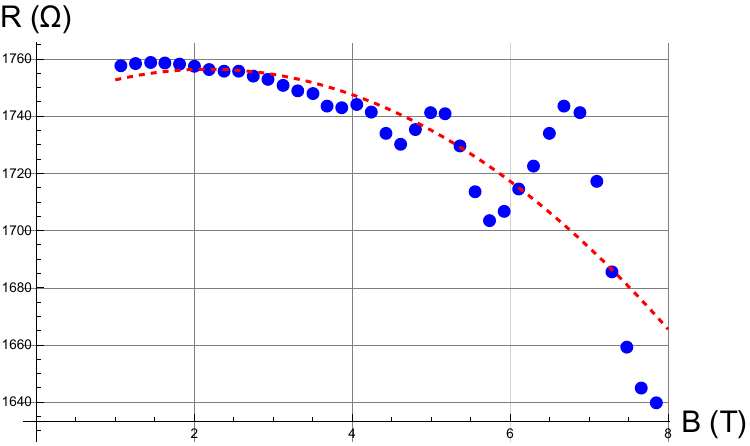}
        \caption{$\theta = 10.3^\circ$}
        \label{fig:angle10.3}
    \end{subfigure}
    \hfill
    \begin{subfigure}[b]{0.24\textwidth}
        \centering
        \includegraphics[width=\linewidth]{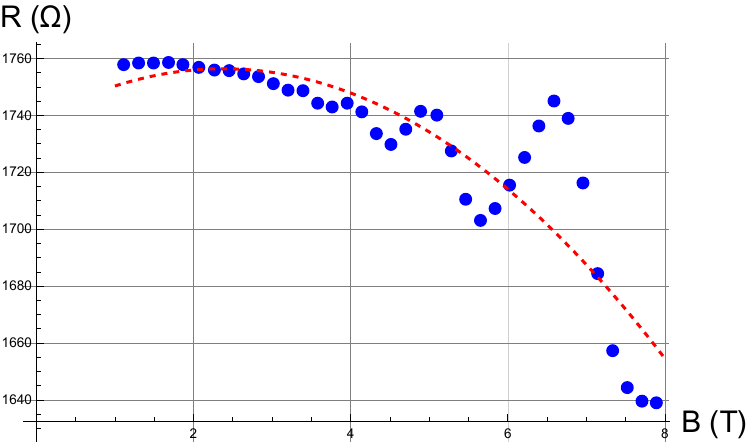}
        \caption{$\theta = 0^\circ$}
        \label{fig:angle0}
    \end{subfigure}
    
    \caption{Comparison of experimental data and fitted curves for different magnetic field orientations. The blue points represent the measured resistance $R$ as a function of magnetic field $B$, while the red dashed lines show the corresponding fits. Angles $\theta$ indicate the orientation of the magnetic field relative to the sample plane.}
    \label{fig:all_fits}
\end{figure}

\section{Other experimental results}

\begin{figure}[ht!]
\centering
\fbox{\includegraphics[trim={70pt 0pt 60pt 0pt}, clip,angle=-90,width=\linewidth]{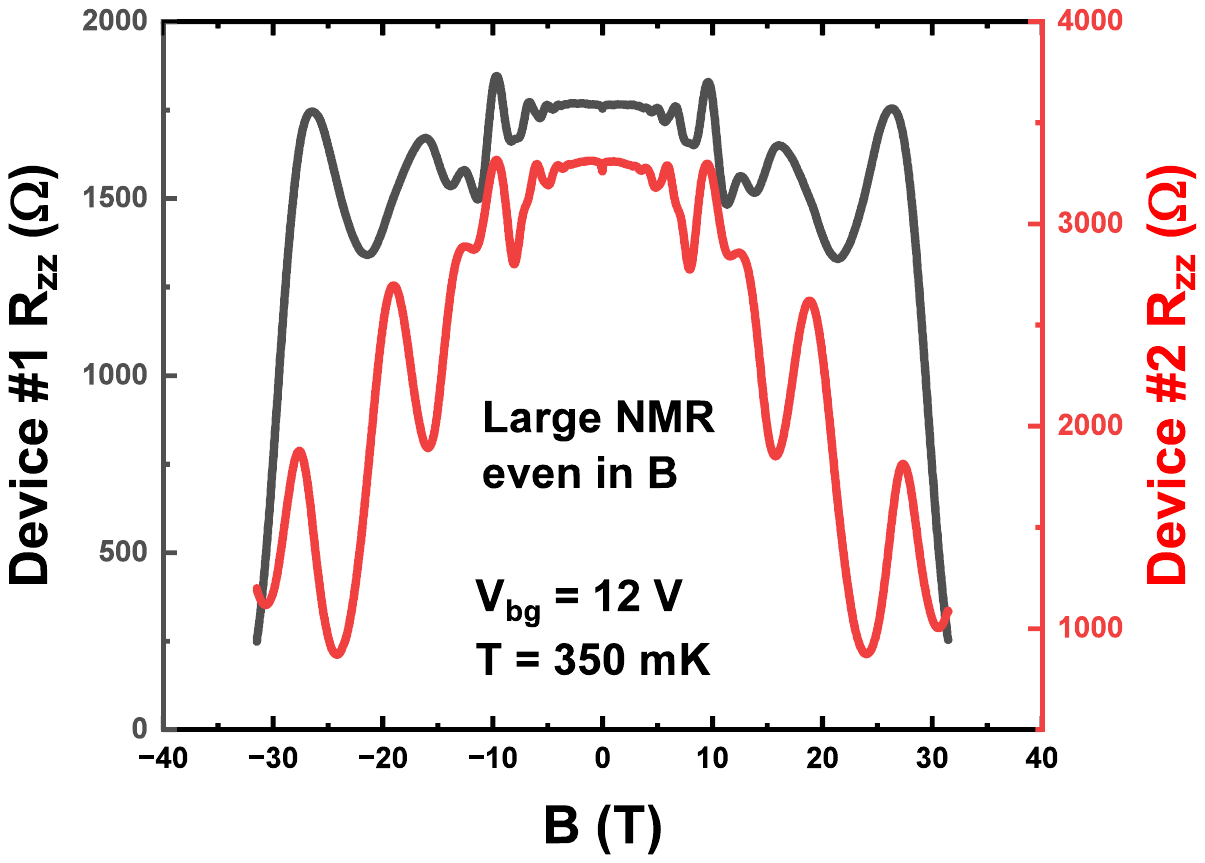}}
\caption{\textbf{GNMR in the conduction band of 2D Te.}
GNMR measured in two different devices with similar carrier densities, exhibiting symmetric negative magnetoresistance from $-30$ T to $+30$ T.}
\label{Fig-Pannel-05}
\end{figure}

\begin{figure}[ht!]
\centering
\fbox{\includegraphics[trim={70pt 0pt 60pt 0pt}, clip,angle=-90,width=\linewidth]{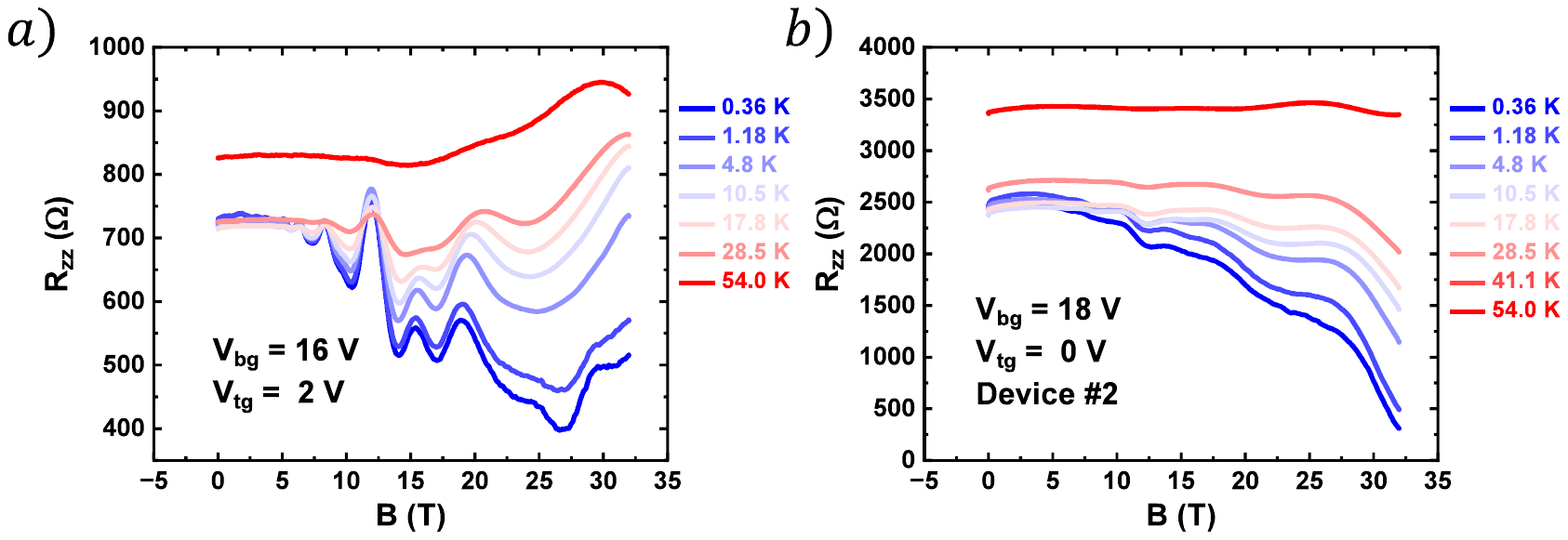}}
\caption{\textbf{Temperature-dependent GNMR in 2D Te.}
(a) Temperature dependence of the negative magnetoresistance at higher carrier densities measured in the same device as in Fig. 3d.
(b) Temperature dependence of the negative magnetoresistance measured in a different device, showing similar behavior with suppression of NMR at 54 K.}
\label{Fig-Pannel-05}
\end{figure}